\documentclass[
	aps, prd, reprint,
	10pt, notitlepage, a4paper,
        floats, floatfix,
	amsmath, amssymb, amsfonts, eqsecnum,
	superscriptaddress,
	showpacs, showkeys,
	nofootinbib,
 	longbibliography,
]{revtex4-1}

\usepackage{aas_macros}
\usepackage[usenames,dvipsnames]{xcolor}
\usepackage{xspace} 
\usepackage{bm,graphicx} 
\usepackage[utf8]{inputenc} 
\usepackage[normalem]{ulem}

\xdefinecolor{mylinkcolor}{rgb}{0,0,0.5}
\usepackage[
	bookmarksnumbered, bookmarksopen, bookmarksopenlevel=2,
	breaklinks=true,
	colorlinks=true, filecolor=mylinkcolor, citecolor=mylinkcolor,
	linkcolor=mylinkcolor, urlcolor=mylinkcolor, menucolor=mylinkcolor,
]{hyperref}

\def\vct#1{{\bm{#1}}}

\def\nl{\\ & \quad}

\def\ee{\end{equation}}
\def\eea{\end{eqnarray}}
\def\be{\begin{equation}}
\def\bea{\begin{eqnarray}}

\DeclareMathOperator{\Order}{\mathcal{O}}
\DeclareMathOperator{\sgn}{sgn}

\def\Lcov{\Lambda}

\allowdisplaybreaks

\begin{document}

\newcommand{\AEI}{\affiliation{Max-Planck-Institute for Gravitational Physics (Albert-Einstein-Institute),
\\ Am M{\"u}hlenberg 1, 14476 Potsdam-Golm, Germany, EU}}
\newcommand{\UU}{\affiliation{Institute for Theoretical Physics,
Utrecht University,\\ Princetonplein 5, 3584 CC Utrecht, The Netherlands}}
\newcommand{\UNH}{\affiliation{Department of Physics, University of New Hampshire, 9 Library Way, Durham NH 03824, USA}}
\newcommand{\Potsdam}{\affiliation{Institut f{\"u}r Physik und Astronomie, Universit{\"a}t Potsdam, Haus 28,
Karl-Liebknecht-Str. 24/25, 14476, Potsdam, Germany}}

\title{Spin effects on neutron star fundamental-mode dynamical tides:\\phenomenology and comparison to numerical simulations}

\date{\today}

\author{Jan Steinhoff}
\AEI
\author{Tanja Hinderer}
\UU
\author{Tim Dietrich}
\Potsdam\AEI
\author{Francois Foucart}
\UNH

\begin{abstract}
 Gravitational waves from neutron star binary inspirals contain information on strongly-interacting matter in unexplored, extreme regimes. Extracting this requires robust theoretical models of the signatures of matter in the gravitational-wave signals due to spin and tidal effects. In fact, spins can have a significant impact on the tidal excitation of the quasi-normal modes of a neutron star, which is not included in current state-of-the-art waveform models. We develop a simple approximate description that accounts for the Coriolis effect of spin on the tidal excitation of the neutron star's quadrupolar and octupolar fundamental quasi-normal modes and incorporate it in the \texttt{SEOBNRv4T} waveform model. We show that the Coriolis effect introduces only one new interaction term in an effective action in the co-rotating frame of the star, and fix the coefficient by considering the spin-induced shift in the resonance frequencies that has been computed numerically for the mode frequencies of rotating neutron stars in the literature. We investigate the impact of relativistic corrections due to the gravitational redshift and frame-dragging effects, and identify important directions where more detailed theoretical developments are needed in the future. Comparisons of our new model to numerical relativity simulations of double neutron star and neutron star-black hole binaries show improved consistency in the agreement compared to current models used in data analysis.
\end{abstract}

\maketitle


\section{Introduction}
The gravitational waves (GWs) from inspiraling binary systems encode detailed information about the nature and internal structure of the compact objects. These signatures arise from spin and tidal effects, including \emph{dynamical tides} associated with the excitation of the objects' characteristic quasi-normal modes. This is particularly interesting for neutron stars (NSs), where gravity compresses matter up to several times the normal nuclear density~\cite{Lattimer:2015nhk,Rezzolla:2018jee}, making NSs unique laboratories for the ground state of strongly interacting matter at the highest physically possible densities. The new opportunities for characterizing such matter with GWs were demonstrated with the first binary NS merger event GW170817 ~\cite{Abbott:2017dke,Abbott:2018wiz,TheLIGOScientific:2017qsa,TheLIGOScientific:2017qsa,Bauswein:2017vtn,Annala:2017llu,Most:2018hfd,Ruiz:2017due,Margalit:2017dij,Rezzolla:2017aly,Shibata:2017xdx,Abbott:2018exr,Abbott:2018wiz,De:2018uhw,Radice:2017lry,Coughlin:2018miv,Coughlin:2018fis,Dai:2018dca,Radice:2018ozg,Lucca:2019ohp,Capano:2019eae,Dietrich:2020efo,Raaijmakers:2019dks}. In the future, higher-precision GW measurements for populations of NS have the potential to advance our understanding of the fundamental physics of strong interactions as well as the emergent multibody phenomena in subatomic matter. Extracting the information on matter from GW signals from binaries is critically predicated on highly accurate theoretical waveform models that link between features in GWs and source parameters~\cite{Cutler:1994ys,Veitch:2014wba,LIGOScientific:2019hgc}. In particular, the waveform models must include all relevant physical effects. This requires a detailed understanding of the behavior of matter in spinning, relativistic objects under nonlinear, dynamical gravity, which is a challenging task.

A significant research effort in the last decades has focused on developing GW models for binary black holes (BHs), which involve only vacuum gravity and are characterized by only their masses and spins~\cite{Carter:1971zc,Hawking:1971vc,Gurlebeck:2015xpa}. There has also been much recent progress on describing the effects of matter in binary inspirals. A number of studies have focused on GW signatures of dynamical tides associated with different NS modes within different approximations~\cite{Bildsten:1992my,1994ApJ...426..688R,Lai:1993ve,Lai:1993di,1994ApJ...426..688R,1977A&A....57..383Z,Willems:2002na,1970A&A.....4..452Z,1978ASSL...68.....K,Kochanek:1992wk,Hansen:2005qv,Mora:2003wt,Kokkotas:1995xe,Flanagan:2007ix,Ferrari:2011as,Damour:1991yw,Shibata:1993qc,Rathore:2002si,Lai:1993di,Vines:2010ca,Vines:2011ud,Chakrabarti:2013xza,Andersson:2017iav,Yu:2016ltf,Tsang:2013mca,Tsang:2011ad,Pnigouras:2019wmt,Pratten:2019sed,Schmidt:2019wrl,Suvorov:2020tmk,Andersson:2019ahb,Pan:2020tht,Ng:2020etb,Pons:2001xs,Gualtieri:2001cm}, rotational multipole moments~\cite{Krishnendu:2017shb,Bohe:2015ana,Marsat:2014xea,Levi:2015msa,Porto:2012as,Porto:2010zg,Levi:2014gsa,Buonanno:2012rv,Levi:2020lfn,Levi:2020uwu,Levi:2019kgk}, gravitomagnetic tidal interactions~\cite{Flanagan:2006sb,Favata:2005da,Landry:2015cva,Pani:2018inf,Landry:2015snx,Banihashemi:2018xfb,Poisson:2020eki,Poisson:2020mdi,Poisson:2020ify,Gupta:2020lnv}, eccentricity~\cite{Gold:2011df,Chirenti:2016xys,Yang:2019kmf,Vick:2019cun}, nonlinear mode couplings~\cite{Xu:2017hqo,Essick:2016tkn,Landry:2015snx}, spin-tidal couplings in the adiabatic limit~\cite{Jimenez-Forteza:2018buh,Abdelsalhin:2018reg,Pani:2015nua,Pani:2015hfa,Endlich:2015mke,Gagnon-Bischoff:2017tnz,Landry:2017piv,Landry:2015zfa,Dietrich:2019kaq,Levi:2020uwu}, and the effects of spins on the tidal response of black holes~\cite{LeTiec:2020spy,LeTiec:2020bos,Chia:2020yla,Goldberger:2020fot} as well as on dynamical tides in NSs in the Newtonian limit~\cite{Ho:1998hq,Ma:2020rak,Lai:1998yc,Lai:1997wh,Ivanov:2015dua,Lai:2006pr,Ma:2020oni}. Recently, effective-field-theory calculations of tidal effects in scattering events have also come into focus~\cite{Kalin:2020mvi,Kalin:2020lmz}, see also Ref.~\cite{Bini:2020flp}, and Refs.~\cite{Cheung:2020sdj,Haddad:2020que,Cheung:2020gbf,Bern:2020uwk} for analogous work based on massive quantum fields or scattering amplitudes.

Substantial further effort has gone into developing state-of-the-art waveform models for data analysis within the so-called phenomenological \texttt{IMRPhenom}~\cite{Ajith:2007qp, Ajith:2007kx,Khan:2015jqa,Husa:2015iqa,Hannam:2013oca,Schmidt:2012rh,Schmidt:2014iyl,Ajith:2009bn,Santamaria:2010yb,Khan:2019kot,Garcia-Quiros:2020qpx,Dietrich:2017aum,Dietrich:2018uni,Pratten:2020fqn,Pratten:2020ceb} and effective-one-body (EOB) \texttt{SEOBNR}/\texttt{TEOBResumS} families~\cite{Buonanno:1998gg,Buonanno:2000ef,Bohe:2016gbl,Babak:2016tgq,Taracchini:2012ig,Taracchini:2013rva,Pan:2013rra,Pan:2013tva,Barausse:2009xi,Barausse:2009aa,Babak:2016tgq,Barausse:2011ys,Pan:2011gk,Cotesta:2018fcv,Ossokine:2020kjp,Nagar:2018plt,Nagar:2018zoe,Damour:2014sva,Damour:2014yha,Damour:2012ky,Damour:2009kr,Damour:2008te,Damour:2008gu,Damour:2007vq,Messina:2018ghh,Nagar:2017jdw,Nagar:2016ayt,Damour:2001tu,Damour:2008qf,Nagar:2011fx,Balmelli:2013zna,Nagar:2019wds,Nagar:2017jdw,Damour:2014yha,Bernuzzi:2012ku,Nagar:2011aa,Nagar:2011fx} (see also the reviews~\cite{Dietrich:2020eud,Damour:2016bks,Damour:2011xga,Buonanno:2014aza,Damour:2012mv,Hannam:2013pra,Damour:2008yg}). These models all include the effects of spin-induced multipole moments and the dominant tidal effects characterized by equation-of-state-dependent tidal deformability (or Love number) coefficients. In previous work, we calculated the effects of dynamical tides from the fundamental ($f$\nobreakdash-) modes and incorporated them in the \texttt{SEOBNR} models~\cite{Hinderer:2016eia,Steinhoff:2016rfi}, leading e.g. to the \texttt{SEOBNRv4T} model. However, our model of dynamical tides had several limitations. For instance, we did not consider the effects of spin on the tidal response of the NS, which is the most prominent effect of spin-matter interactions.

In this paper, we extend the parameter space of waveform models by accounting for these spin effects in an approximate way, and investigate their role in NSNS and NSBH binaries. As expected on physical grounds and confirmed in previous work, e.g.~\cite{Ho:1998hq,Foucart:2018lhe,Ma:2020rak}, the effect of spins on dynamical $f$-mode tides can significantly enhance the matter signatures in GWs in the late inspiral for anti-aligned spins, depending also on the parameters. We show in this paper that this can lead to non-negligible dephasings with current data analysis models which neglect this effect. It is therefore urgent to model dynamical tides of rotating NS to enable the robustness of using GWs as probes for subatomic physics as the LIGO~\cite{TheLIGOScientific:2014jea}, Virgo~\cite{TheVirgo:2014hva}, and KAGRA~\cite{Akutsu:2020his} GW detectors are improving in sensitivity and new, third-generation facilities are being envisioned. We study three main effects that influence the orbital frequency in a binary in which a rotating NS's $f$-mode is resonantly excited by the tidal field of the companion: (i) the gravitational redshift of the NS, (ii) the relativistic dragging of the NS's inertial frame, including also the additional effects of the orbiting companion, and (iii) the Coriolis effect due to the NS's spin. We derive an estimate for the resonant orbital frequency which approximately takes into account all of these effects and demonstrate that the most important effect is due to the NS's spin because of near-cancellations between the redshift and frame-dragging effects. We develop a simple modification of the existing $f$-mode EOB waveform model which includes the Coriolis effect and is based on introducing spin-dependent shifts in the $f$-mode frequency and tidal deformability coefficients. We test our model against results from numerical relativity simulations both for aligned and anti-aligned spins and find improved consistency compared to current models used in data analysis. Our simple model can readily be used to improve GW measurements. A more detailed theoretical study and model development which also overcomes other limitations and includes dynamical tides in the odd-parity sector will be the subject of forthcoming works. 

The organization of this paper is as follows. We begin in Sec.~\ref{sec:Newtaction} by deriving a Newtonian action for quadrupolar, parity-even dynamical tides in rotating stars. We start from a description in terms of the normal modes for the fluid displacement due to the perturbations and convert to a basis of symmetric-tracefree tensors. That basis is more convenient for identifying selection rules, and for generalizing to relativistic stars. Such a generalization is worked out in Sec.~\ref{sec:relaction} in the co-rotating frame where the background (unperturbed) fluid is at rest, provided we allow for general coupling coefficients not restricted to their Newtonian values. We specialize to the case of $f$-modes, which have the largest tidal couplings, and work to linear order in the rotation frequency. This leads to an effective action with one as yet undetermined coefficient characterizing the Coriolis interaction between the star's spin and its tidal spin, i.e., the angular momentum associated with the dynamical quadrupole. In Sec.~\ref{sec:PN} we extend the action to a binary system and
derive explicit equations of motion within the post-Newtonian approximation for the orbital dynamics. From the solutions for the quadrupole we obtain the response function whose features we analyze in Sec.~\ref{sec:response}. We discuss how to determine the spin-tidal Coriolis coefficient by matching to results for the $f$-mode frequencies of rotating neutron stars from the literature and obtain a quasi-universal relation for this shift. Next, we consider the impact of relativistic effects---gravitational redshift and frame-dragging---on the dynamical tides and quantify their importance. In Sec.~\ref{sec:EOBmap} we derive a simple phenomenological model that accounts for the Coriolis effect by applying spin-dependent shifts of the $f$-mode frequency and tidal deformability parameter in the existing \texttt{SEOBNRv4T} waveform model. We test this model against numerical relativity simulations of spinning binary neutron star and neutron star -- black hole binaries from the BAM and SXS codes in Sec.~\ref{sec:comparisons}. Section~\ref{sec:conclusion} contains our conclusions, and the Appendix contains a brief discussion of the relation of this paper to Ref.~\cite{Ma:2020rak}.

The notation here follows that in Ref.~\cite{Steinhoff:2016rfi}.
We use geometric units with $G=c=1$ throughout.
Capitalized Latin indices $A$, $B$, \dots on tensors denote the representation in the spatial co-rotating frame and take values 1, 2, 3.
Greek letters $\mu$, $\nu$, \dots denote spacetime coordinate indices and run through 0, 1, 2, 3.
Lower-case Latin indices $i$, $j$, \dots run through 1, 2, 3 and denote either spatial coordinate indices when used on position variables or indices in a local euclidean frame comoving with the center of the star for other tensors (spin, quadrupole)~\cite{Levi:2015msa,Steinhoff:2016rfi}.
Boldface notation for vectors with such indices is also used. Round brackets around indices denote the symmetrization,
square brackets denote the corresponding antisymmetric combination, and angle brackets denote symmetric-tracefree projection.
Our convention for the Riemann tensor is
\begin{equation}
R^{\mu}{}_{\nu\alpha\beta} = \Gamma^{\mu}{}_{\nu \beta , \alpha}
        - \Gamma^{\mu}{}_{\nu \alpha , \beta}
        + \Gamma^{\rho}{}_{\nu \beta} \Gamma^{\mu}{}_{\rho \alpha}
        - \Gamma^{\rho}{}_{\nu \alpha} \Gamma^{\mu}{}_{\rho \beta} ,
\end{equation}
where $\Gamma^{\mu}{}_{\nu \beta}$ is the Christoffel symbol.
In the derivations we consider the case with only one extended body, which we label as body 1 with mass $m_1$. Since we work in the regime of linearized tides, the case of two stars can be obtained by adding the same contribution with the body labels exchanged. For a binary system, we define the total mass
$M = m_1 + m_2$, the reduced mass $\mu = m_1 m_2 / M$, and the symmetric mass ratio $\nu = \mu / M$.

\section{Newtonian dynamical tides of rotating stars}
\label{sec:Newtaction}
In this section we recapitulate Newtonian tides as linear perturbations
of a background solution for a star in equilibrium following Refs.~\cite{1964ApJ...139..664C,Steinhoff:2016rfi,Chakrabarti:2013xza,Flanagan:2007ix,Gupta:2020lnv} (see also, e.g., Refs.~\cite{Schenk:2001zm,1967MNRAS.136..293L,Rathore:2002si}).
The perturbation is
described by a displacement vector field of the fluid elements
$\vct\xi(\vct x, t)$ away from their background position.
It is useful to consider the function space of all displacements as a
complex Hilbert space with an inner product
\begin{equation}
  \langle \vct{\xi}, \vct{\xi}' \rangle = \int d^3 x\rho_0 \, \vct{\xi}^* \! \cdot \vct{\xi}' , \label{inner}
\end{equation}
where $\rho_0$ is the unperturbed mass density of the background configuration.

We restrict the discussion here to an ideal fluid with barotropic equation of
state $\rho(p)$ relating the mass density $\rho$ and the isotropic pressure $p$. That is, we neglect effects from, e.g., temperature, viscosity, and
buoyancy, which play a subdominant role for the fundamental modes in neutron stars. We will first briefly recall the non-rotating case and obtain the Lagrangian describing the dynamics of the tidal perturbations, then generalize to include effects of spin to linear order in the rotation frequency, and finally transform to a description in terms of symmetric-tracefree tensors.

\subsection{Nonrotating stars}
\label{sec:Newtnonrot}
We first consider a nonrotating, hence spherically symmetric,
star in equilibrium with density $ \rho_0(r)$. The star is then placed in an external gravitational potential $\Phi$, which induces dynamical perturbations to the fluid. We will consider the perturbations only to linear order. For instance, the mass density perturbation is $\delta \! \rho = - \nabla (\rho_0 \vct \xi)$, where $\vct{\xi} = \vct{\xi}^*$ is the physical fluid displacement. 
The equations of motion for the dynamical tidal
perturbations can be derived from a Lagrangian for the fluid displacement given by
\begin{equation}
\label{eq:LDTfluidnospin}
  L_\text{DT} = \frac{1}{2} \langle \dot{\vct{\xi}}, \dot{\vct{\xi}} \rangle - \frac{1}{2} \langle \vct{\xi}, \mathcal{D} \vct{\xi} \rangle + \langle \vct{f}^{\rm ext}, \vct \xi \rangle .
\end{equation}
The first term in Eq.~\eqref{eq:LDTfluidnospin} is the kinetic energy of the perturbation, the second term specifies the energies associated with the internal restoring forces, while the last term
is the potential energy in the external field. For the case considered in this paper, the external force\footnote{We disregard here the fictitious force arising from the center-of-mass acceleration of the star (see, e.g., \cite{Steinhoff:2016rfi}), which effectively just cancels $\vct{f}^\text{ext}$ in the dipolar sector (equivalence principle).} is $\vct{f}^{\rm ext}=-\nabla \Phi$ where $\Phi$ is the gravitational potential of a binary companion orbiting at a distance $\vct{r}(t)$ and given by 
\begin{equation}
\Phi=-\frac{m_2}{|\vct{x}-\vct{r}(t)|} .
\label{eq:binaryPhi}
\end{equation}
The linear operator $\mathcal D$ is defined by
\begin{equation}
\label{eq:Doperator}
  \mathcal{D} \vct \xi = - \nabla \left\{ \left[ \frac{c_s^2}{\rho_0} + 4 \pi  \Delta^{-1} \right] \nabla \cdot (\rho_0 \vct \xi) \right\} ,
\end{equation}
where $c_s=\sqrt{(\partial p_0/\partial \rho_0)}$ is the speed of sound of the background fluid configuration with pressure $p_0$. The first term in Eq.~\eqref{eq:Doperator} comes from the perturbation of the internal energy, and the second (nonlocal) term describes the gravitational self-energy of the perturbation.

The operator $\mathcal{D}$ is Hermitian with respect to the inner product (\ref{inner}).
Thus its eigenvectors $\vct \xi_{n\ell m}$ are an orthonormal basis of the normal modes labeled by the type of mode $n$, the multipolar order $\ell$, and an angular-momentum number $m$ associated 
with a decomposition into (vector) spherical harmonics. The real eigenvalues $\omega_{n\ell}^2$, where $\omega_{n\ell}$ is the mode frequency, are determined from
\begin{equation}
  \mathcal{D} \vct \xi_{n\ell m} = \omega_{n\ell}^2 \vct \xi_{n\ell m} , \quad
  \langle \vct \xi_{n\ell m} , \vct \xi_{n'\ell' m'} \rangle = \delta_{nn'} \delta_{\ell\ell'} \delta_{mm'} .
\end{equation}
The mode frequencies $\omega_{n\ell}$ are degenerate over $m$ because the operator $\mathcal{D}$ is rotation symmetric.  Similarly, due to parity invariance of $\mathcal{D}$,
the modes can be categorized as even parity (electric-type) or odd-parity (magnetic-type).
As the integration measure $\rho_0 d^3 x$ of the inner product (\ref{inner})
has compact support, the normal modes $\vct{\xi}_{n\ell m}$ are countable and are
enumerated by the number $n$ (besides $\ell$ and $m$).
We restrict our attention to pressure modes in this paper and take $n$ to be the number of radial nodes.
The fundamental pressure mode or $f$-mode is then labeled by $n=0$.

We can decompose any fluid displacement $\vct \xi$ into the orthonormal basis
of the normal modes,
\begin{equation}
  \vct \xi = \sum_{n\ell m} q_{n\ell m}(t) \vct \xi_{n\ell m}(\vct x), \quad
  q_{n\ell m} = \langle \vct \xi_{n\ell m}, \vct \xi \rangle , \label{modedecomp}
\end{equation}
with time-dependent amplitudes $q_{n\ell m}(t)$.
The reality condition $\vct{\xi} = \vct{\xi}^*$ implies that
$q_{n\ell m} ^* = (-1)^m q_{n\ell\, -m}$, which follows from the analogous relation for
the spherical harmonics. The general Lagrangian then reads
\begin{equation}
\label{eq:LDT1nospin}
  L_\text{DT} = \sum_{n\ell m} \left[ \frac{1}{2} |\dot{q}_{n\ell m}|^2 - \frac{1}{2} \omega_{n\ell}^2 |q_{n\ell m}|^2 + \langle \vct{f}^{\rm ext}, \vct \xi_{n\ell m} \rangle \, q_{n\ell m} \right] .
\end{equation}
To compute the coefficients for the overlap between the external field and the mode functions, and to identify the modes giving the most important contributions to $L_{\rm DT}$, it is useful to express the potential from Eq.~\eqref{eq:binaryPhi} as a Taylor series expansion around the center of the star. Choosing coordinates such that the center of the star is located at $\vct{x}=0$, the expansion of the potential is
\begin{equation}
\Phi(t, \vct{x})=\Phi(t, \vct{0}) +x^j \partial_j \Phi(t, \vct{x})|_{\vct{x}=0}+\sum_{\ell=2}^\infty \frac{1}{\ell!} x^L  E_L
\end{equation}
where the $\ell$-th tidal moments $E_L$ for $\ell\geq 2$ are defined by (following the conventions in \cite{Steinhoff:2016rfi}):
\begin{equation}
\label{eq:Eijdef}
 E_L=\partial_L \Phi \! \mid_{\vct{x}=0} ,
\end{equation}
and $L=A,B,C, \ldots$ denotes a string of $\ell$ indices.
Note that $E_L$ is symmetric and tracefree, which follows from $\Delta \Phi \! \mid_{\vct{x}=0} = 0$.
The tidal potential can equivalently be written as a spherical harmonic multipolar expansion given by
\begin{equation}
\label{eq:sphericalharmonicpotential}
\Phi=-m_2\sum_{\ell,m} \frac{4\pi} {2\ell+1}\frac{|\vct{x}|^\ell}{r(t)^{\ell+1}}Y_{\ell m}\left(\frac{\pi}{2}, \phi\right)Y_{\ell m}^*(\theta, \varphi),
\end{equation}
where $\vct{x}$ and $(\theta, \varphi)$ are associated with a comoving coordinate system centered on the star, and $r, \phi$ characterize the orbital coordinates in the equatorial plane. We define the overlap integral $I_{n\ell}$ by
\begin{equation}\label{Inl}
I_{n\ell}=\langle \nabla |\vct{x}|^\ell Y_{\ell m}(\theta, \varphi), \vct \xi_{n\ell m} \rangle .
\end{equation}
The term $\langle \vct{f}^{\rm ext}, \vct \xi_{n\ell m} \rangle$ in the Lagrangian can then be written as
\begin{equation}
\langle \vct{f}^{\rm ext}, \vct \xi_{n\ell m} \rangle=-\frac{N_\ell}{\ell!}I_{n\ell}E^*_{\ell m}, 
\end{equation}
where
\begin{equation}\label{eq:Elm}
  E_{\ell m}=N_\ell \mathcal{Y}^{*\ell m}_{L}E_L = - (2\ell-1)!!\, N_\ell Y^*_{\ell m}\left(\frac{\pi}{2}, \phi\right) \frac{m_2}{r^{\ell+1}}.
\end{equation}
This can be either obtained from Eq.~\eqref{eq:Eijdef} or its spherical-harmonic analog Eq.~\eqref{eq:sphericalharmonicpotential}. Recall that one can convert between spherical harmonics and unit vectors using the identity $Y_{\ell m}=\mathcal{Y}^{\ell m}_{L}n^{\langle L\rangle}$ and defining the coefficient $N_\ell = \sqrt{4 \pi \ell! / (2 \ell + 1)!!}$ that arises when applying the inverse conversion to change from $n^{\langle A}n^{B\rangle}$ to $Y_{\ell m}$; see e.g.~Ref.~\cite{Gupta:2020lnv} for useful formulas.

The modes with the largest contributions to $L_{\rm DT}$ can be identified by the following considerations. First, we note that the $\ell$th multipole of the external tidal field associated with $\ell\geq 2$ derivatives of $\Phi$ is increasingly suppressed for increasing
multipole orders $\ell$. We thus expect the dominant contributions to come from the low-$\ell$ modes. However, the $\ell=0,1$, as well as all magnetic-type modes do not couple linearly to the external
gravitational field in the Newtonian case. The $\ell=0$ interaction is forbidden due to the conservation of mass,
while the $\ell=1$ interaction leads to an overall motion of the star, which has no gauge-invariant physical meaning according to the
weak equivalence principle (universality of free fall). The magnetic modes
couple linearly to gravitomagnetic tidal fields which is a relativistic phenomenon that is absent in Newtonian gravity.
Hence, to leading order, the external field drives the electric quadrupolar
($\ell=2$) modes, so we restrict our attention to them in the following. 

\subsubsection{Transformation to the basis of symmetric-tracefree Cartesian tensors}
We can equivalently express the Lagrangian in terms of symmetric-tracefree tensors
using the conversion between spherical harmonics and unit vectors provided by the symmetric-tracefree tensors ${\cal Y}^{\ell m}_L$. The mode amplitudes $q_{n\ell m}$ can then be directly translated to Cartesian tensors. For the quadrupole $\ell=2$, we adopt the normalization
\begin{equation}
Q_n^{AB} = \sqrt{2 \lambda_n} \omega_n N_2 \sum_m \mathcal{Y}_{AB}^{2m} q_{n2m}, 
\end{equation}
where $N_2 = \sqrt{8\pi / 15}$, $\omega_n = \omega_{n2}$ is the mode frequency, and $\lambda_n$ is the tidal deformability of the mode, related here to the overlap integral by\footnote{Our convention for $I_{n\ell}$ differs from Refs.~\cite{Chakrabarti:2013xza,Steinhoff:2016rfi} by a factor of $N_\ell$, see Eq.~(2.2) in Ref.~\cite{Steinhoff:2016rfi}.}
\begin{equation}
\lambda_n= \frac{N_2^2 I_{n2}^2}{2\omega_n^2} = \frac{4\pi I_{n2}^2}{15 \omega_n^2} .
\end{equation}
The total quadrupole is given by summing over all overtones
\begin{equation}
Q_{AB}=\sum_{n=0}^\infty Q_n^{AB}.
\end{equation}
We also define the Newtonian quadrupolar tidal tensor 
\begin{equation}
E^{AB} = \partial_{A} \partial_{B } \Phi(\vct{x})\mid_{\vct{x}=0}.
\end{equation}
The Lagrangian~\eqref{eq:LDT1nospin} can then be written as
\begin{multline}
  L_\text{DT} = \sum_n \bigg[ \frac{1}{4 \lambda_n \omega_n^2} \left( \dot Q_n^{AB} \dot Q_n^{AB}  - \omega_n^2 Q_n^{AB} Q_n^{AB} \right) \\ - \frac{1}{2} E^{AB} Q_n^{AB} \bigg] .
\end{multline}
We remind the reader that $Q_n^{AB}$ is the contribution of the $n$-mode to the (symmetric-tracefree)
quadrupole of the star, and $E^{AB}$ is the external tidal field evaluated at
the center of the star. A key point to note is that because we work in a 3-dimensional rest-frame of the star labeled by $A, B = 1,2,3$ (or the corotating frame below), the structure of the couplings for the internal dynamics of the quadrupole is the same for Newtonian and relativistic stars, cf.~Eq.~(1.4) in \cite{Steinhoff:2016rfi}; the distinction between them is only through the coefficients ($\lambda_n$, $\omega_n$). We will exploit this fact for rotating stars below, since we are interested here in fully relativistic NSs, where $\omega_n$ and $\lambda_n$ are computed in general relativity. 


\subsection{Rotating stars}\label{Newtonrot}
It is straightforward to extend the discussion from the last section
to stars that are rotating uniformly with an angular velocity of the star $\vct \Omega$ as observed in the inertial frame. It is convenient
to describe the star in the corotating frame, where the background fluid
elements are at rest.  At linear order in $\vct \Omega$, the only new interaction with dynamical multipoles is due to the
Coriolis force,
\begin{equation}
  L_\text{DT} = \frac{1}{2} \langle \dot{\vct{\xi}}, \dot{\vct{\xi}} \rangle
  - \langle \vct \xi, \vct \Omega \times \dot{\vct{\xi}} \rangle
  - \frac{1}{2} \langle \vct{\xi}, \mathcal{D} \vct{\xi} \rangle
 + \langle \vct{f}^{\rm ext}, \vct \xi \rangle .
\end{equation}
The background star gets deformed away from spherical
symmetry only at quadratic order in $\vct \Omega$, so that
the eigenvectors and -values of $\mathcal{D}$ are approximately the same as for a spherically-symmetric
nonrotating star. Inserting the decomposition for $\vct \xi$ (\ref{modedecomp}) leads to
\begin{multline}
  L_\text{DT} = \sum_{n\ell m} \bigg[ \frac{1}{2} |\dot{q}_{n\ell m}|^2 - \frac{1}{2} \omega_{n\ell}^2 |q_{n\ell m}|^2
+\frac{1}{\ell!} I_{n\ell}E_{\ell m} \, q_{n\ell m} \\
    - \sum_{n'\ell' m'} q_{n\ell m}^* \dot{q}_{n'\ell' m'} \langle \vct \xi_{n\ell m}, \vct \Omega \times \vct \xi_{n'\ell' m'} \rangle \bigg] .
  \end{multline}
The last term here represents a linear mode coupling (quadratic in the action)
due to the rotation. These mode couplings are subject to selection rules. The selection rules are most readily identified in the symmetric-tracefree basis where they are automatically implemented when imposing symmetry requirements. Specifically, the allowed couplings are all parity-invariant contractions between the symmetric-tracefree tensors of the modes with either the parity-odd angular velocity vector $\vct \Omega$ or its associated antisymmetric parity-even tensor 
\begin{equation}
\Omega^{AB} = \epsilon_{ABC} \Omega^C.
\end{equation}
We focus here on the electric quadrupolar ($\ell=2$) modes $Q_n^{AB}$.
All possible spin-mode couplings involving $Q_n^{AB}$ read
\begin{equation}
Q_n^{AB} \dot Q_{n'}^{BC} \Omega^{AC}, \quad
Q_n^{AB} \dot Q_{n'}^B \Omega^A, \quad
Q_n^{AB} \dot Q_{n'}^{ABC} \Omega^C .
\end{equation}
Hence, couplings to magnetic modes are possible to a dipole $Q_{n'}^{B}$ and to
an octupole $Q_{n'}^{ABC}$, which we neglect since these magnetic modes
are not externally driven in the Newtonian limit. We then arrive at the Lagrangian
\begin{equation}
\label{eq:LDTNewt}
  L_\text{DT} \approx \sum_n \bigg[ L_o - \frac{1}{2} E^{AB} Q_n^{AB} +  L_{\rm SQ}\bigg],
\end{equation}
where the oscillator and spin-mode contributions are given by
\begin{align}
L_o=&\frac{1}{4 \lambda_{n} \omega_{n}^2} \left( \dot Q_n^{AB} \dot Q_n^{AB}  - \omega_{n}^2 Q_n^{AB} Q_n^{AB} \right) , \\
L_{\rm SQ}=&\sum_{n'} C_{\Omega n n'} \Omega^{AB} Q_n^{AC} \dot{Q}_{n'}^{CB}.
\end{align}
The effect of rotation is explicit here via the term $L_{\rm SQ}$ describing a rotation-induced coupling between different modes of the same multipolar order due to the Coriolis force. We have inserted the as-yet-undetermined coefficients $C_{\Omega n n^\prime}$.
It is possible to write down an explicit formula for $C_{\Omega n n^\prime}$, analogous to Eq.~\eqref{Inl} for $I_{n\ell}$, that is valid in Newtonian gravity. However, we do not need this here since we are ultimately interested in the fully relativistic value of this coefficient.
We will focus here on the fundamental $f$-modes with $n=n^\prime=0$ and determine all coefficients $\lambda_0$, $\omega_0$, $C_{\Omega 00}$ by matching to relativistic results for the effect of spin on the mode frequencies in Sec.~\ref{sec:matching} below.
We will drop the label $n$ on $Q$ from now on.

\section{Relativistic dynamical tides of rotating stars}
\label{sec:relaction}

In this section, we upgrade the Newtonian action to a relativistic one, following the nonrotating case in Ref.~\cite{Steinhoff:2016rfi}.
For the treatment of the star's spin or angular momentum, we draw from Refs.~\cite{Porto:2005ac,Goldberger:2009qd,Levi:2010zu,Levi:2015msa,Endlich:2015mke}.
The resulting action is an effective one, where length scales below the bodies' size are integrated out, and could also be constructed from an effective-field-theory approach~\cite{Goldberger:2004jt,Goldberger:2007hy,Foffa:2013qca,Rothstein:2014sra,Porto:2016pyg,Levi:2018nxp}.
However, we do not attempt here a rigorous construction of such an action based on symmetries and power-counting arguments.
Instead, we only include terms in the relativistic action that are already present in the Newtonian case above, but with undetermined coefficients.
We expect terms that are absent in the Newtonian limit to be suppressed for relativistic electric tides, which is also justified from numerical studies of the relativistic tidal response~\cite{Chakrabarti:2013lua}.
We note, however,  that in the case of magnetic tides, such an approach would crucially miss important terms in the relativistic effective action, as discussed in~\cite{Gupta:2020lnv}.
The explicit results of calculations in the post-Newtonian approximation for the binary dynamics based on effective actions can be found in, e.g., in Refs.~\cite{Bini:2012gu,Henry:2019xhg,Henry:2020ski,Henry:2020pzq}.

\subsection{Upgrading the Newtonian action}

A relativistic rotating star can be represented by a worldline $y^\mu(\tau)$ with dynamical tidal and spin degrees of freedom propagating along it.
Here $\tau$ is the proper time and the tangent 4-velocity is given by $U^\mu = d y^\mu(\tau) / d \tau$ such that $U_\mu U^\mu = -1$.
The rotation of the star can be encoded by an orthonormal corotating, body-fixed frame $\Lambda_I{}^\mu(\tau)$ on the worldline (with $I, J, \dots = 1,2,3$ and $\Lambda_I{}^\mu \Lambda_{J\mu} = \delta_{IJ}$) describing the orientation of the star.
This frame is taken to be comoving such that $\Lambda_I{}^\mu U_\mu = 0$. 
Based on this frame, we can define the relativistic angular velocity in the corotating frame as
\begin{equation}
  \Omega^I = \epsilon^{IJK} \Omega_{JK}, \qquad \Omega_{JK} = \frac{D \Lambda_I{}^\mu}{d \tau} \Lambda_{J\mu} ,
\end{equation}
where $D$ is the covariant differential. 
The external tidal field is given by the electric part of the Weyl tensor as $E_{IJ} = \Lcov_I{}^\mu U^\alpha \Lcov_J{}^\nu U^\beta C_{\mu\alpha\nu\beta}$ in the relativistic case, where $C_{\mu\alpha\nu\beta}$ is the Weyl curvature tensor.
The relativistic $f$\nobreakdash-mode amplitude $Q^{AB}(\tau) \equiv Q_0^{AB}(\tau)$ is written in the corotating frame along the worldline. Time derivatives are taken with respect to proper time and denoted by an overdot $\dot{~}=d/d\tau$.

It is now straightforward to upgrade the above Newtonian Lagrangian~\eqref{eq:LDTNewt} to the relativistic case: the structure remains the same but all the quantities must be computed from the appropriate relativistic definitions discussed above.
For the description of a binary, one needs to supplement the action $\mathcal{S}$ by a nontidal (NT) part,
\begin{align}
  \mathcal{S} &= \int d\tau \, (\underbrace{L_\text{NT} + L_\text{DT}}_{\displaystyle L}), \\
  L_\text{NT} &= - m_0 + \frac{I}{2} \Omega_I \Omega^I + \dots ,
\end{align}
with the irreducible (rotation-independent) mass $m_0$, the moment of inertia $I$, and the dots representing further terms not relevant here (e.g., from the spin-induced quadrupole moment~\cite{1967ApJ...150.1005H,Poisson:1997ha,Laarakkers:1997hb}).

\subsection{Legendre transformation}

The spin-tidal interaction due to the Coriolis force specialized to $n=n^\prime=0$ can also be expressed as a coupling between the star's spin and the tidal spin associated with the $f$-modes as we will show next. This highlights the connection to spin or frame-dragging effects in general relativity. Following Ref.~\cite{Steinhoff:2016rfi}, we define a tidal spin tensor and vector associated with the quadrupolar $f$-modes by
\begin{equation}
   S_{Q}^{A} = \frac{1}{2}\epsilon^A{}_{BC} S^{BC}_{Q}, \qquad S^{AB}_{Q}=4 Q^{C[A}P^{B]}{}_{C},
\end{equation}
where 
\begin{equation}
P_{AB}=\frac{\partial L}{\partial \dot Q^{AB}} ,
\end{equation}
is the conjugate momentum to $\dot Q_{AB}$. 
We also introduce the total spin $S^I_t$ conjugate to the rotation frequency
\begin{equation}
S^I_t= \frac{\partial L}{\partial \Omega_I} ,
\end{equation}
and the associated spin tensor $S_t^{IJ}=\epsilon^{IJ}{}_K S_t^K$.
To zeroth order in the tidal contributions $S_t^J \approx I \Omega^J$.

We perform a Legendre transformation of the Lagrangian to $P_{AB}$ and $S_t^A$, which leads to the action
\begin{align}
\label{eq:relaction_corot}
  \mathcal{S} &= \int d\tau \, (S_t^A \Omega_A + P_{AB} \dot Q^{AB} + R) , \\
  \begin{split}
    R &\approx - m_0 - \frac{1}{2I} S_{tA} S_t^A + C_{SQ} S_{tA} S_Q^A - \frac{1}{2} E_{AB} Q^{AB} \nl
    - \frac{1}{4\lambda_0\omega_0^2} \left( 4 \lambda_0^2 \omega_0^4 P_{AB} P^{AB} + \omega_0^2 Q_{AB} Q^{AB} \right) ,
  \end{split}
\end{align}
where we neglected terms beyond quadratic order in the tidal variables and defined 
\begin{equation}\label{eq:cSQdef}
  C_{SQ} = C_{\Omega 00} \frac{\lambda_0 \omega_{0}^2}{I} .
\end{equation}
We see that the leading-order spin-tidal interaction can be understood as a spin-spin interaction between the ordinary and tidal spins, $S_{tA} S_Q^A $.
We note that due to the dynamical tidal interactions, the spin length $S_t=\sqrt{S_{tA} S_t^A}$ is not constant.

\subsection{Coordinate-frame action}

Let us now connect the relativistic action~\eqref{eq:relaction_corot} to that of the nonspinning case discussed in Ref.~\cite{Steinhoff:2016rfi}, which is formulated in the coordinate frame instead of the corotating frame. The transformation of the above results to the coordinate frame is simply given by   $Q^{AB} = \Lambda^{A\mu} \Lambda^{B\nu} Q_{\mu\nu}$, and similarly for the other tensors. This leads to the relation
\begin{equation}
P_{AB} \dot Q^{AB} = P_{\mu\nu} \frac{D Q^{\mu\nu}}{d \tau} - \Omega_{\mu} S_Q^{\mu} .
\end{equation}
It is convenient to absorb the second term by splitting the total spin $S_t^\mu$ into a ``rotational-only'' spin $S^\mu$ and the tidal part $S_Q^\mu$,
\begin{equation}
  S_t^{\mu} = S^{\mu} + S_Q^{\mu} .
\end{equation}
The coordinate-frame action then reads
\begin{align}\label{eq:Seff}
  \mathcal{S} &= \int d\tau \, \left[ S^\mu \Omega_\mu + P_{\mu\nu} \frac{D Q^{\mu\nu}}{d\tau} + R \right] , \\
  \begin{split}
    R &\approx - m(S^2) + \bar C_{SQ} S_{\mu} S_Q^\mu - \frac{1}{2} E_{\mu\nu} Q^{\mu\nu} \nl
    - \frac{1}{4\lambda_0\omega_0^2} \left( 4 \lambda_0^2 \omega_0^4 P_{\mu\nu} P^{\mu\nu} + \omega_0^2 Q_{\mu\nu} Q^{\mu\nu} \right) ,
  \end{split}
\end{align}
where the coordinate-frame coefficient associated with the spin-tidal interaction is given by
\begin{equation}\label{eq:cbardef}
  \bar C_{SQ} = C_{SQ} - \frac{1}{I} . 
\end{equation}
We are going to find below that this equation just encodes the difference of the $f$-mode frequency between corotating and inertial frames.
We have introduced the constant ADM mass (not to be confused with the magnetic number $m$),
\begin{equation}
  m(S^2) = m_0 + \frac{1}{2 I} S^2 + \Order(S^4) .
\end{equation}
Note that the spin-length is now constant, $S^2 = S_{\mu} S^\mu = \text{const}$. 
Furthermore, since $\Lambda_I{}^\mu U_\mu = 0$, all coordinate-frame tensors are orthogonal to the 4-velocity,
\begin{align}
  \begin{split}\label{eq:varconst}
  \Omega_\mu U^\mu = 0 , & \qquad Q^{\mu\nu} U_\nu =0 , \\
  S^\mu U_\mu = 0 , & \qquad P_{\mu\nu} U^\nu =0 .
  \end{split}
\end{align}
These constraints have to be fulfilled alongside the variational principle for the action.

\section{Post-Newtonian approximation}
\label{sec:PN}

The action above~\eqref{eq:Seff} models a single neutron star interacting with an external gravitational field as a worldline (point-particle) effective action with spin and dynamical quadrupole moment.
Based on this building block, the action of a binary system can be constructed as two copies of Eq.~\eqref{eq:Seff} together with the Einstein-Hilbert action for the gravitational field.
One can further eliminate (integrate out) the orbital-scale field within the post-Newtonian approximation, which is a weak-field and slow-motion approximation around the Newtonian limit.
This can be understood as a formal expansion in the inverse speed of light.
We discuss this post-Newtonian approximate action for a binary system in this section.

\subsection{Post-Newtonian action and Hamiltonian}

Instead of performing the post-Newtonian calculation in detail, we can take a shortcut by building on previous results.
Indeed, the worldline action~\eqref{eq:Seff} is a sum of the nonspinning dynamical tidal action from Ref.~\cite{Steinhoff:2016rfi} and the spin action from, e.g., Ref.~\cite{Levi:2015msa}, plus the simple spin-tidal correction $\bar C_{SQ} S_{\mu} S_Q^\mu$.
Since we work to linear order in spin and tidal interactions, the result for the Lagrangian of a binary system in the post-Newtonian approximation can be taken from these references and by adding the spin-tidal interaction for each body. This leads to
\begin{equation}
  \mathcal{S}_{\rm SQ} = \int d\tau \, \bar C_{SQ} \vct{S}_{Q} \cdot \vct{S}
  = \int dt \, z \, \bar C_{SQ} \vct{S}_{Q} \cdot \vct{S} ,
\end{equation}
where $z = d\tau / dt= \sqrt{g_{\mu\nu} \dot y^\mu \dot y^\nu}$ is the redshift variable. In this section, the meaning of an overdot changes compared to Sec.~\ref{sec:relaction} and now denotes a derivative with respect to coordinate time, $\dot ~ = d / dt$.
In the post-Newtonian results, the temporal components of spin and tidal variables are eliminated by writing them in a comoving local euclidean frame and using Eq.~\eqref{eq:varconst}, see Refs.~\cite{Levi:2015msa,Steinhoff:2016rfi} for details.
The spatial components in this frame are simply denoted as, e.g., $Q^{ij}$ where $i,j = 1,2,3$.

We assume that only one of the objects has a finite size and label it as body 1. Within our approximations, the case of two extended objects can readily be obtained by adding the same tidal contributions with the appropriate parameters for the other object. The action for the binary in the center-of-mass frame in Hamiltonian form has the structure
\begin{equation}
  \label{eq:SPN}
    \mathcal{S}_\text{PN} = \int dt \big( \vct{p} \cdot \vct{r} + \vct{S}_1 \cdot \vct{\Omega}_1 + \vct{S}_2 \cdot \vct{\Omega}_2 + P^{ij} \dot{Q}^{ij} - H_\text{PN} \big) .
\end{equation}
Here, $\vct{p}$ is the relative linear momentum and $\vct{r}$ is the separation vector. The post-Newtonian Hamiltonian splits as $H_\text{PN} = H_\text{NT} + H_\text{DT}$ into a non-tidal $H_\text{NT}$ and a dynamical-tidal $H_\text{DT}$ part with~\cite{Steinhoff:2016rfi}
\begin{equation}
\label{eq:HDTrelativistic}
  H_\text{DT} = z H_o + \vct{\Omega}_\text{FD} \cdot \vct{S}_{Q} - z \bar C_{SQ} \vct{S}_1 \cdot \vct{S}_{Q} + \frac{z}{2} E_{ij} Q^{ij} .
\end{equation}
Different versions of the non-tidal Hamiltonian $H_\text{NT} = H_\text{NT}(\vct{r}, \vct{p}, \vct{S}_1, \vct{S}_2; m_1, m_2, C_W)$ exist in the literature, e.g., in Refs.~\cite{Khalil:2020mmr,Levi:2016ofk}; the precise version will not be important here. 
Here $C_W$ collectively denotes several Wilson coefficients in the original worldline effective action that describe, e.g., spin-induced multipole moments.
Furthermore, the redshift $z$ and frame-dragging/spin-precession frequency $\vct{\Omega}_\text{FD}$ in the Hamiltonian~\eqref{eq:HDTrelativistic} are given by
\begin{equation}
   z = \frac{\partial H_{\rm NT}}{\partial m_1} , \quad
  \vct{\Omega}_\text{FD} = \left. \frac{\partial H_{\rm NT}}{\partial \vct{S}_1} \right\rvert_{C_W=0} ,
\end{equation}
and the oscillator-part Hamiltonian reads
\begin{equation}
  H_o = \frac{1}{4\lambda_0\omega_0^2} \left( 4 \lambda_0^2 \omega_0^4 P^{ij} P^{ij} + \omega_0^2 Q^{ij} Q^{ij} \right) .
\end{equation}
Finally, the post-Newtonian tidal field $E_{ij}=E_{ij}(\vct{r}, \vct{p}; m_1, m_2)$ in the comoving local euclidean frame~\cite{Steinhoff:2016rfi} can be found in Refs.~\cite{Bini:2012gu,Steinhoff:2016rfi} in different gauges.
To leading Newtonian order, the tidal field follows from Eq.~\eqref{eq:Eijdef},
\begin{equation}
  E_{ij} = - \frac{3 m_2}{r^3} n^{\langle i} n^{j \rangle} + \dots ,
\end{equation}
where $r = |\vct{r}|$ and $\vct{n} = \vct{r} / r$;
the explicit form of higher-order corrections will be irrelevant for our studies below.

\subsection{Circular-orbit tidal equations of motion}

From now on, we assume that the binary is on a circular orbit and that the spins are not precessing, i.e. they are aligned or anti-aligned with the orbital angular momentum.
For generic orbits, different gauge choices for $H_\text{NT}$ lead to different expressions for $z$ and $\vct{\Omega}_\text{FD}$.
For circular orbits and nonprecessing spins, however, they are universally given by
\begin{align}
  z &= 1 + \frac{x}{2} (\nu - 3 X_2 ) \nl
      + \frac{x^2}{24} ( 5 \nu^2 - 9 \nu - 6 \nu X_2 - 27 X_2 ) + \Order(x^{5/2}) , \nonumber \\
  \label{eq:FD}
  \frac{\Omega_\text{FD}}{\omega_\text{orb}} &= \frac{x}{2} (\nu + 3 X_2) - x^{3/2} X_2^2 \chi_2 \nl
    - \frac{x^2}{24} ( \nu^2 - 45 \nu + 30 \nu X_2 - 27 X_2) + \Order(x^{5/2}), \nonumber
\end{align}
where $\Omega_\text{FD} = |\vct{\Omega}_\text{FD}|$. We have also defined the spin magnitude $\chi_2 = \pm |\vct{S}_2| / m_2^2$ and the mass ratio $X_2 = m_2 / M$ for the companion, and introduced the frequency variable  $x = (M \omega_\text{orb})^{2/3}$ where $\omega_\text{orb}$ is the orbital frequency. For aligned companion spin, $\chi_2 >0$ while for anti-aligned spin it is $\chi_2<0$.
Here the power on $x$ corresponds to the post-Newtonian order.
Note that for circular orbits, the binary is in equilibrium, so that $\omega_\text{orb}$, $z$, and $\Omega_\text{FD}$ are constant.

An interesting observation here is that the frame dragging due to the orbital angular momentum given by the first term in~\eqref{eq:FD} and that resulting from the companion spin, given by the second term in~\eqref{eq:FD}, have opposite signs.
This can be understood by visualizing the directed gravitomagnetic field lines analogous to a bar magnet (see, e.g., the discussion in Ref.~\cite{Schaefer:2004qh}).
The neutron star experiences the gravitomagnetic field from the orbital motion at its source (``inside the gravito-magnet''), where the field lines point in the same direction as the orbital angular momentum. Conversely, the field of the companion felt by the star is outside its source, where the gravito-magnetic field lines point in the opposite direction of $\vct{S}_2$. This implies that for aligned companion spins, the net frame-dragging effects are smaller than for anti-aligned spins of the companion. 

Hamilton's equations of motion follow from varying the tidal variables in the action~\eqref{eq:SPN},
\begin{equation}
  \dot{Q}^{ij} = \frac{\partial H_\text{DT}}{\partial P^{ij}} , \qquad
  \dot{P}^{ij} = - \frac{\partial H_\text{DT}}{\partial Q^{ij}} ,
\end{equation}
or more explicitly
\begin{align}
  \dot{Q}^{ij} &= 2 z \lambda_0 \omega_0^2 P^{ij} + 2 \Omega_\text{FD}^{k(i} Q^{j)k} - 2 z \bar C_{SQ} S_1^{k(i} Q^{j)k}  , \\
  \dot{P}^{ij} &= - \frac{z}{2 \lambda_0} Q^{ij} + 2 \Omega_\text{FD}^{k(i} P^{j)k} - 2 z \bar C_{SQ} S_1^{k(i} P^{j)k} - \frac{z}{2} E_{ij} .
\end{align}
These equations can be decoupled by transforming to the spherical-harmonic $(\ell, m)$ basis.
For this purpose, we express the relativistic quadrupole due to the $f$-modes as $Q^{ij}= N_2 \sum_k \mathcal{Y}_{AB}^{2m} Q_m$, and similar for $P^{ij}$ and $E_{ij}$, analogous to the transformation in the Newtonian case discussed in Sec.~\ref{sec:Newtnonrot}.
The reality condition implies that $Q_m^*=(-1)^m Q_{-m}$.
Furthermore, we choose to align the z-axis with the spin so that $(S_1^i) = S_1 (0, 0, 1)$. Since we also assume that the spins are collinear with the orbital angular momentum, this means that $\vct{\Omega}_\text{FD}$ is along the z-axis as well.
This leads us to the equations of motion
\begin{align}  
  \dot{Q}_m &= 2 z \lambda_0 \omega_0^2 P_m - i m (\Omega_\text{FD} - z \bar C_{SQ} S_1) Q_m , \\
  \dot{P}_m &= - \frac{z}{2 \lambda_0} Q_m - i m (\Omega_\text{FD} - z \bar C_{SQ} S_1) P_m  - \frac{z}{2} E_m .
\end{align}
These equations can be combined into a single second-order differential equation describing the $f$-mode oscillations,
\begin{equation}
  \label{eq:TEOMlm}
  [ \partial_t + i m (\Omega_\text{FD} - z \bar C_{SQ} S_1) ]^2 Q_m + z^2 \omega_0^2 Q_m = - z^2 \omega_0^2 \lambda_0 E_m .
\end{equation}
For the driving force at Newtonian order it holds $E_m = \mathcal{E}_m e^{-im \omega_\text{orb} t}$ with
\begin{equation}\label{eq:force}
  \mathcal{E}_{\pm 2} = -\sqrt{\frac{3}{2}} \mathcal{E}_0 = - \frac{3 m_2}{2 M^3} x^3 , \quad
  \mathcal{E}_{\pm 1}= 0 .
\end{equation}
Note that in the Newtonian limit $x \approx M / r$.
This can be obtained from Eq.~\eqref{eq:Elm} noting that $E_m = E_{2m}$ and $\phi = \omega_\text{orb} t$.
The latter relation is consistent with our assumption of an equilibrium solution, but only holds approximately for an inspiral and breaks down close to the resonance, where $\omega_\text{orb}$ changes in time.
This will be discussed further below in connection with the effective Love number.

\section{Exploring the tidal response}
\label{sec:response} 

In this section, we derive the frequency-domain response of the quadrupole $Q^{ij}$ to the tidal field $E_{ij}$ in a binary system described by the post-Newtonian Hamiltonian $H_\text{DT}$. We emphasize again that we are treating $Q^{ij}$ as fully relativistic, and only use the post-Newtonian approximation for quantities related to the orbital dynamics. We assume again a circular-orbit nonprecessing binary.
Our goal is to investigate the impact of the redshift, frame-dragging, and spin-tidal coupling on the resonance frequency.
We accomplish these aims by considering the response function, the frequency-dependent ratio of the induced quadrupole to the tidal field, which encodes these effects. Having calculated the response function, we first match the spin-tidal coupling constant to numerical results for the $f$-mode frequency of isolated spinning neutron stars. We then find a quasi-universal relation for this coupling, i.e., a relation that is approximately independent of the equation of state for the nuclear matter. Subsequently we identify relativistic effects due to redshift and frame dragging on the resonance frequency in a binary and quantify their importance. 

\subsection{The tidal response function}

To compute the response function, it is easiest to work in the Fourier domain. We take the Fourier transform, denoted by a tilde, of the dynamical tidal variables according to the conventions
\begin{equation}
Q_m(t)=\int \frac{d \omega}{2\pi} \tilde Q_m(\omega) e^{-i \omega t} ,
\end{equation}
and similarly for $E_m$.
It is now straightforward to solve the equation of motion in spherical-harmonic basis~\eqref{eq:TEOMlm} for $\tilde Q_m(\omega)$,
\begin{equation}\label{eq:Qsol}
  \tilde Q_m = - \tilde F_m \tilde E_m
\end{equation}
with $\tilde{E}_m = 2\pi \mathcal{E}_m \delta(\omega - m \omega_\text{orb})$. We find that the gravitoelectric quadrupolar frequency-domain tidal response is given by
\begin{equation}
  \label{eq:response}
  \tilde{F}_m = \frac{z^2 \omega_0^2 \lambda_0}{z^2 \omega_0^2 - [ \omega - m \Omega_\text{FD} + m z \bar C_{SQ} S_1 ]^2} .
\end{equation}
For the limiting case of adiabatic tides $\omega \sim m \omega_\text{orb} \rightarrow 0$ the response function reduces to the tidal deformability $\tilde{F}_m(\omega = 0) =  \lambda_0$, noticing that $\Omega_\text{FD}(x=0) = 0$ and that we neglect terms quadratic in $S_1$.
Hence, the response function is a generalization of the Love number $\lambda_0$ to dynamical frequency-dependent tides.
The poles of the response correspond to a resonant excitation of the $f$-mode. We will exploit this fact in our analyses below.

At this point it is convenient to pick a sign convention for the frequencies.
For our purpose, it is most useful to assume a fixed sign of the orbital driving frequency as $\omega_\text{orb}>0$ and allow for both signs for the spin $S_1$ encoding its orientation (aligned $S_1>0$ or antialigned $S_1<0$).
Since $\omega = m \omega_\text{orb}$ it follows that $\omega \lessgtr 0$ for $m \lessgtr 0$, or $|\omega| = \sgn(m) \omega$.
Prograde/retrograde motion of the tidal bulge (relative to the neutron star spin) corresponds to $\sgn(m) = \pm \sgn(S_1)$.

\subsection{Matching the spin-tidal coupling}\label{sec:matching}

To determine the spin-tidal coupling coefficient $\bar C_{SQ}$, it is sufficient to consider an isolated neutron star without a companion. For such a star the redshift reduces to $z=1$ and the frame-dragging vanishes $\Omega_\text{FD} = 0$.
The poles of the response~\eqref{eq:response} are located at frequencies $\omega$ equal to the inertial-frame $f$-mode frequency $\omega_f > 0$ of the spinning neutron star, i.e., at $|\omega| = \omega_f$.
This leads us to the identification of the spin-induced shift of the $f$-mode frequency
\begin{equation}
  \label{eq:deltaomega0}
  \Delta \omega_0 \equiv \omega_f - \omega_0= - |m| \bar C_{SQ} S_1 , 
\end{equation}
where $\omega_0$ is the $f$-mode frequency of a nonspinning star. In terms of the constant $C_{SQ}$ in Eq.~\eqref{eq:cbardef} it holds $\Delta \omega_0 - |m| \Omega_1 = - |m| C_{SQ} S_1$, which is the corotating-frame frequency shift; the corotating-frame $f$-mode frequency reads $\tilde \omega_f = \omega_f - |m| \Omega_1$ and $S_1 = I \Omega_1$. In the remainder of this section we will drop the label $1$ on $\Omega$; the meaning that it is the rotation frequency of the extended body will be implied. 

To fix the relativistic value of the spin-tidal coupling $\bar C_{SQ}$, we compare the effective frequency shift from Eq.~\eqref{eq:deltaomega0} to results for the $f$-mode frequencies of rotating relativistic NSs in the Cowling approximation from Ref.~\cite{Doneva:2013zqa} (see also \cite{Gaertig:2008uz,Gaertig:2010kc}), specializing to the slow-rotation regime.
Specifically, in Ref.~\cite{Doneva:2013zqa}, Doneva et al provide quadratic ploynomial fits for the $|m| = \ell$ $f$-mode frequencies in the corotating frame for the stable (s) and unstable (u) branches of the form
\begin{equation}\label{shiftfit}
\frac{\tilde{\omega}_{\ell}}{\omega_0}=1+a^{s/u}_\ell \left| \frac{\Omega}{\Omega_K} \right|-b^{s/u}_\ell \left| \frac{\Omega}{\Omega_K} \right|^2.
\end{equation}
Note that the $\tilde{\omega}_{\ell}$ are defined to be positive, just like our $\omega_f$.
In our convention, the stable/prograde branch corresponds to $\Omega > 0$ and the unstable/retrograde one to $\Omega < 0$.
The coefficients were determined in Eqs.~(21)--(24) of Ref.~\cite{Doneva:2013zqa} to be $a^{u}_2=0.402$, $b^u_2=-0.406$, $a^{u}_3=0.373$, $b^u_2=-0.485$, and $a^{s}=-0.235$, $b^s=-0.358$ for both $\ell=2,3$.
The parameter $\Omega_K$, the Kepler frequency, was found to be well-approximated by $\Omega_K\left[{\rm kHz}\right]=2\pi\left[1.716\sqrt{\bar \rho_0}-0.189\right]$, where $\bar\rho_0=(m_1/1.4M_\odot)/(R/10{\rm km})^3$ is the scaled mean density of the non-rotating background solution with mass $m_1$ and radius $R$.
Using the transformation to inertial-frame frequencies $\omega_\ell = \tilde{\omega}_\ell + |m| \Omega$ (which actually flips the sign of the frequency shift), we find the spin-induced shift of the frequency $\Delta \omega_0 = \omega_{\ell=2} - \omega_0$ and hence a matching for the spin-tidal coupling (recalling $|m| = \ell = 2$),
\begin{equation}\label{eq:match}
  \bar C_{SQ} = \frac{\omega_0 - \tilde{\omega}_{\ell=2} - 2 \Omega}{2 I \Omega} .
\end{equation}
To find a constant value for $\bar C_{SQ}$, one should take here the limit of small rotation frequency $\Omega_1 \equiv \Omega \rightarrow 0$.
We note that we obtain $\omega_0$ using our own code in this paper, solving the linear perturbation equations of nonrotating neutron stars (without making use of the Cowling approximation).
A quasi-universal fit for $\Delta \omega_0$ is also given in Eq.~(4) of Ref.~\cite{Kruger:2019zuz}, which does not make use of the Cowling approximation, but is restricted so far to $\ell=2$. 

\subsection{Universality of the coupling}

We find that the spin-tidal coupling fulfills a quasi-universal relation that is approximately independent of the equation-of-state and given by
\begin{equation}\label{universalshift}
\bar C_{SQ} \approx - \frac{3}{4I} , \quad \text{or} \quad \Delta \omega_0\approx \frac{3}{2} \Omega .
\end{equation}
For the purpose of checking this relation, we calculate $\Omega_K$ using the \texttt{RNS} code \cite{Stergioulas:Morsink:1999,Stergioulas:1994ea} and $\omega_0$ from the perturbation equations of nonrotating NSs (specifically the version given in Ref.~\cite{Chakrabarti:2013lua}) for the MS1b and SLy equations of state and neutron star masses ranging between $1.1$--$2.0M_\odot$.
Inserting these values for $\Omega_K$ and $\omega_0$ into above fit~\eqref{shiftfit} leads to $\bar C_{SQ}$ via Eq.~\eqref{eq:match}.
Note that the stable and unstabe branches are described by different signs of $\Omega$ here and are averaged over to arrive at Eq.~\eqref{universalshift}.
This symmetry between stable and unstable branches in the linear regime, which can be inferred from Eq.~\eqref{eq:deltaomega0} here, is not manifest in the fit~\eqref{shiftfit}, which is based on data points that are mostly in the regime nonlinear in $\Omega$.
Clearly, it would be desirable to check Eq.~\eqref{universalshift}, which describes the linear regime, within a slow rotation approximation in the future (and without making use of the Cowling approximation).

Note that Eq.~\eqref{universalshift} is consistent with the Newtonian case considered in Ref.~\cite{1996A&A...311..155L}.
We also checked our relation against the updated fit in Eq.~(4) of Ref.~\cite{Kruger:2019zuz}, which leads to $\Delta \omega_0 \approx 2 \pi |a_1^{s/u}| \Omega$ and it holds $2\pi |a_1^{s/u}| = 1.2 \dots 1.4$ which is slightly lower than our rough estimate of $3/2$.
But still the fit in Ref.~\cite{Kruger:2019zuz} might not be optimal in the slow-rotation regime (i.e., most data points are for fast rotation).
However, Ref.~\cite{Kruger:2019zuz} provides an optimal extension of the quasi-universal relation above to fast rotation for $\ell=2$.

Finally, we also consider the shift in the octopole $(\ell=3)$ sector, which we estimate through a similar procedure as for the quadrupole explained above. We find that the octopole $f$-mode frequency $\omega_{03}$ is effectively shifted by 
\begin{equation}
\Delta \omega_{03}\approx \frac{5}{2} \Omega .
\end{equation}

In order to obtain definite values for the frequency shift given the spin $S = I \Omega$, we also need to know the moment of inertia $I$.
For neutron stars, the moment of inertia is related to the dimensionless Love numbers $\Lambda=\lambda_0/m_1^5$ through a nearly universal relation (that holds over a wide range of equations of state) of the form
\begin{equation}
\ln(I)\approx\sum_{i=0}^4 c_i (\ln\Lambda)^i
\end{equation}
where the coefficients are given in Table I of Ref.~\cite{Yagi:2016bkt} as $c_0=1.496$, $c_1=0.05951$, $c_2=0.02238$, $c_3=-6.953\times 10^{-4}$, and $c_4=8.345\times 10^{-6}$. 

\subsection{Relativistic effects on the resonance frequency}\label{sec:relativistic}

Let us now investigate the $f$-mode resonances in a binary system.
As for the spin effects, we will use the poles in $\omega$ of the response function~\eqref{eq:response} to determine these effects.
The driving force $\tilde E_m$ can only excite a resonance for $|m| = 2$.
Using $\omega = m \omega_\text{orb}$, this leads to the resonance condition
\begin{equation}
  \label{eq:omegares}
  \omega_\text{orb} - \Omega_\text{FD} = z \frac{\omega_f}{2} , \quad \text{(at resonance)}
\end{equation}
We can interpret this in the following way:
The frame-dragging frequency effectively shifts the orbital frequency that the neutron star experiences, while the redshift factor effectively reduces the mode frequency.

\begin{figure}
  \includegraphics[width=0.48\textwidth]{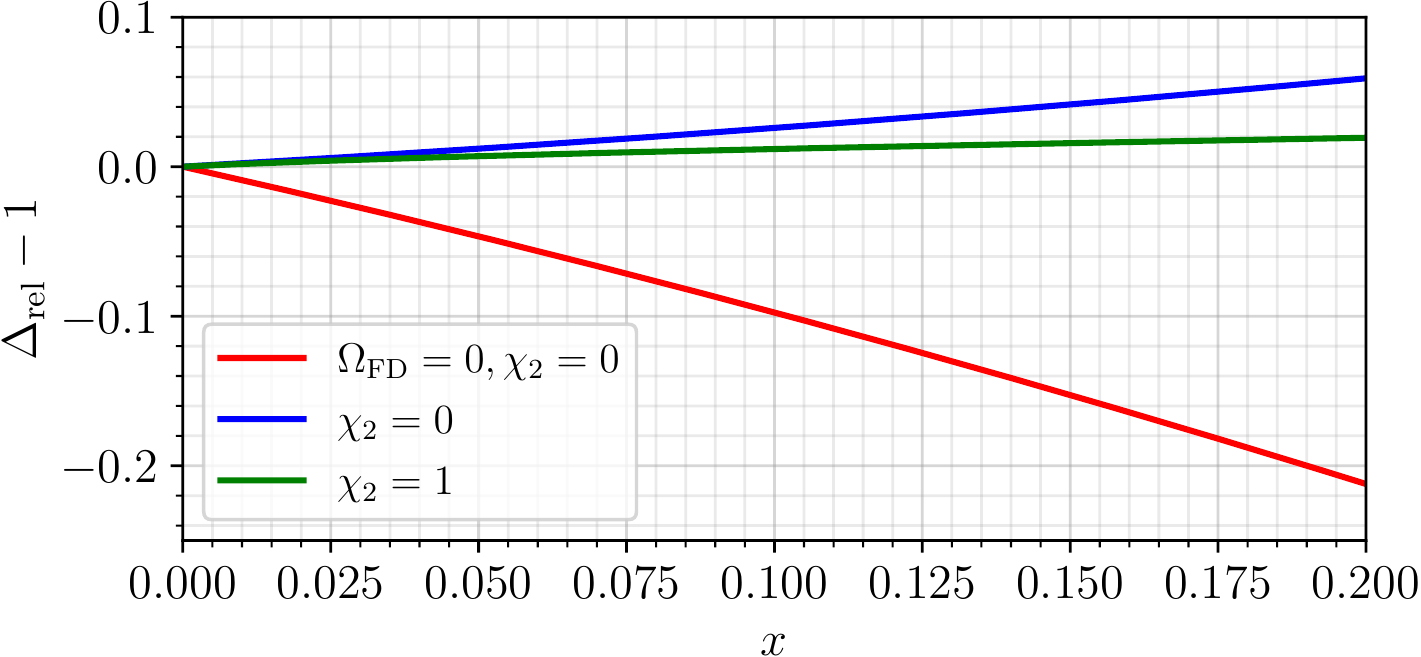}
  \caption{\emph{Relativistic effects on the resonance frequency~\eqref{eq:relshift}} for mass ratio $q = m_2 / m_1 = 2$.
    The red curve shows only the redshift effect, the blue curve neglects the companion spin, and the green curve is with ``maximal'' aligned spin on the companion.
    The effect of the black-hole spin becomes larger with increasing mass ratio, and is opposite for anti-aligned spin.}
   \label{fig:relshift}
\end{figure}

In the absence of relativistic redshift and frame-dragging effects, the resonance happens at an orbital frequency of $\omega_f / 2$.
Thus, it makes sense to normalize the relativistic resonant frequency $\omega_\text{orb} = \omega_\text{orb}^\text{res}$ from Eq.~\eqref{eq:omegares} as
\begin{align}\label{eq:relshift}
  &\Delta_\text{rel} \equiv \frac{2 \omega_\text{orb}^\text{res}}{\omega_f} = \frac{z}{1 - \Omega_\text{FD} / \omega_\text{orb}} \\
  &= 1 + \nu x - \chi_2 X_2^2 x^{3/2} + \frac{9+4\nu}{6} \nu x^2 + \Order(x^{5/2}) ,
\end{align}
such that $\Delta_\text{rel} \approx 1$ at the resonance in the absence of relativistic effects.
The result for the relativistic shifts of the resonance in terms of $\Delta_\text{rel}$ are displayed in Fig.~\ref{fig:relshift}; see Sec.~I.B of Ref.~\cite{Steinhoff:2016rfi} for a detailed interpretation.

We see in Fig.~\ref{fig:relshift} that the redshift (red curve) and frame dragging effects almost cancel out (blue curve almost at $\Delta_\text{rel} \approx 1$) for comparable-mass binaries.
This was already noted qualitatively in Ref.~\cite{Steinhoff:2016rfi}, and is now quantified by Eq.~\eqref{eq:relshift}.
We note that numerical simulations of eccentric binaries, e.g.~\cite{Chaurasia:2018zhg}, found that the radiation emanating from the neutron-star oscillations shows only the redshift but no noticeable frame-dragging effects.
This is not immediately in conflict with our observation, which considers the orbital frequency (and radiation produced by the orbital motion), but it would be desirable to investigate relativistic effects on the radiation emanating from the neutron-star oscillations analytically in future work.

The frame dragging generated by the companion spin also shifts the resonance frequency, but this effect is small for comparable-mass binaries (see the discussion above regarding the sign of this dragging).
This changes with increasing mass of the companion, such that the companion spin can dominate over the orbital angular momentum.
For larger mass black hole companions, however, the net effect of tidal interactions decreases and becomes more difficult to discern. 
To conclude, for a large part of the binary parameter space relevant for neutron stars, we can approximately neglect the relativistic effects on the resonance, $\Delta_\text{rel} \approx 1$. They would be important for broader applications to black hole mimickers and waveform models for third-generation detectors, which is outside the scope of this paper. For neutron star binaries, the dominant effect on the resonance is due to the spin-tidal coupling~\eqref{eq:deltaomega0}.
We hence proceed in the next section with a Newtonian approximation and incorporate the spin-tidal interaction in the effective Love number introduced in Ref.~\cite{Hinderer:2016eia,Steinhoff:2016rfi}.

\section{Adapting the \texttt{SEOBNRv4T} model}
\label{sec:EOBmap}

In the preceding section, we identified the spin-tidal coupling and the corresponding shift of the tidal-resonance frequency as the most important spin effect on dynamical tides.
In this section, we incorporate this spin-tidal coupling in the \texttt{SEOBNRv4T} model. We have implemented these modifications in the LIGO Algorithms Library \texttt{LALSuite} at: \url{https://github.com/jsteinhoff/lalsuite/tree/tidal_resonance_NSspin}. In this model, the dynamical $f$-mode tidal effects are included through an effective Love number, calculated in the Newtonian limit, that approximately captures the frequency-dependence of the response~\cite{Hinderer:2016eia,Steinhoff:2016rfi}.
The model is still relativistic since it utilizes post-Newtonian results for the tidal interaction $\sim E_{\mu\nu} E^{\mu\nu}$, currently to next-to-next-to leading order~\cite{Bini:2012gu,Henry:2019xhg,Henry:2020ski,Henry:2020pzq}.
The effective Love number is calculated from an analysis of the solution for the oscillator amplitudes $Q_m$ before and during the $f$-mode resonance. Below, we discuss the main modifications to this due to the Coriolis effect. 
A comparison to related work in Ref.~\cite{Ma:2020rak} is given in Appendix~\ref{sec:compareMa}.

We first consider the solutions for $Q_m$ before the resonance.\footnote{Our previous work~\cite{Hinderer:2016eia,Steinhoff:2016rfi} used a different notation for quadrupole components, given by
\begin{equation*}
(Q^{ij}) = \begin{pmatrix} \alpha+b & c & 0\\
       c & \alpha-b &  0\\
 0& 0& -2\alpha
       \end{pmatrix} .
\end{equation*}
These variables are related to the $(2, m)$ degrees of freedom by $\alpha=- Q_{0} / \sqrt{6}$,
$b=(Q_{2}+Q_{-2}) / 2 = \Re(Q_2)$, and $c=i(Q_{2}-Q_{-2}) / 2=-\Im(Q_2)$.}
We note that the $m=0$ mode has a vanishing frequency at linear order in spin and cannot be resonantly excited.
Furthermore, for $|m| = 1$ the driving force~\eqref{eq:force} vanishes. Thus, the only contributions to the resonance are associated with $|m| = 2$.
Gathering the pre-resonance solution (where $\omega_\text{orb} \approx \text{const}$) for $Q_{2}=Q^*_{-2}$ from Eqs.~\eqref{eq:Qsol}, \eqref{eq:force}, and \eqref{eq:response}, neglecting relativistic effects from redshift $z\approx 1$ and frame dragging $\Omega_\text{FD} \approx 0$ (as justified in the previous section), and transforming back to the time domain leads to
\begin{equation}
Q_2=\frac{-\lambda_0 \omega_0^2{\cal E}_2 e^{-2i\omega_\text{orb} t}}{\omega_0^2-(2\omega_{\rm orb}-\Delta\omega_0)^2}, \label{eq:bcouter}
\end{equation}
with $\Delta \omega_0=-2 \bar C_{SQ} S_1$, see Eq.~\eqref{eq:deltaomega0}. Note that although we are assuming that $\Delta \omega/\omega_0$ is small, we do not expand the denominator in the solution. This is important to preserve the underlying physics of the resonance shift, and is a common approach for oscillators with small perturbations to their equations of motion (e.g., for an anharmonic oscillator~\cite{Landau1976Mechanics}). 

Near the resonance, the denominator of the solutions~\eqref{eq:bcouter} vanishes and the dynamics require a local analysis that accounts for the evolution of $\omega_{\rm orb}$ due to gravitational radiation. This was discussed for the nonspinning case in \cite{Hinderer:2016eia,Steinhoff:2016rfi} in terms of two-timescale expansions that exploit the hierarchy between the timescales in the system associated with the orbital motion $\sim \omega_{\rm orb}^{-1}$, the $f$-mode oscillations $\sim \omega_0^{-1}$, and the gravitational radiation reaction $\tilde t=\epsilon \phi$, where $\phi=\int \omega_{\rm orb}dt$ and $\epsilon=256\mu M^{2/3}(\omega_f/2)^{5/3}/5$ is a small dimensionless parameter; the temporal width of the resonance is intermediate between the orbital and radiation reaction timescales. Here, we promote these nonspinning results to the spinning case with minor but important modifications. We will obtain approximate results of the dominant effect without re-doing the entire calculations by using different physical perspectives of the Coriolis effect for the analysis away from and near the resonance, as we now discuss. The asymptotic behavior of the solution \eqref{eq:bcouter} near the resonance is
\begin{equation}
\label{eq:outernearres}
\lim_{\omega_{\rm orb}\to \frac{\omega_0+\Delta \omega_0}{2}}Q_2=-\frac{\lambda_0 \bar{\cal E} e^{2i( \tilde t_{\rm res}/\epsilon+\hat t/\sqrt{\epsilon}  )}  }{2\, \sqrt{\epsilon}\, \hat t \, | \bar\Omega^\prime|}+O(\hat t^{-3}, \epsilon^0),
\end{equation}
Here, $\hat t=\sqrt{\epsilon}(\phi-\phi_{\rm res})$ is a rescaled shifted phase variable and '$\rm res$' denotes evaluation at the resonance. The quantity $\bar{\cal E}=- {\cal E}_2\omega_0^2/(2\omega_{\rm orb}-\Delta\omega_0)^2$ is a rescaled tidal amplitude. The function $\bar\Omega$ is the frequency ratio between the $f$-mode and tidal driving frequencies. Its rescaled derivative evaluated at the resonance is given by
\begin{eqnarray}
\bar \Omega^\prime&=& \left. \frac{d}{d\tilde t}\left(\frac{\omega_0}{2\omega_{\rm orb}-\Delta \omega_0}\right)\right\rvert_{\rm res}\nonumber\\
&=&\frac{\dot\omega_{\rm orb}}{\epsilon \omega_{\rm orb}}\frac{\partial}{\partial \omega_{\rm orb}}\left[\frac{\omega_0}{2\omega_{\rm orb}-\Delta \omega_0}\right]\bigg\rvert_{\omega_{\rm orb}=(\omega_0+\Delta \omega_0)/2}
\end{eqnarray}
In the nonspinning case when $\Delta \omega_0=0$, this expression evaluates to be $\bar \Omega^\prime=-3/8$ using the leading order frequency evolution due to gravitational radiation reaction $\dot \omega_{\rm orb}=96/5\mu M^{2/3}\omega_{\rm orb}^{11/3}$. In the spinning case, however, there is an extra contribution that depends on the frequency shift and we obtain
\begin{equation}
\bar \Omega^\prime=-\frac{3}{8}\frac{\omega_0+\Delta \omega_0}{\omega_0}.
\end{equation}
Next, we consider the solutions in the resonance region. In this regime, the Coriolis effect can be viewed as an effective shift in the $f$-mode frequency. This means that all the results from \cite{Steinhoff:2016rfi} carry over in a straightforward manner with the only change being a shift in $\omega_f$. The inner solutions are thus given by
\begin{eqnarray}
Q_2^{\rm res}
&=& \frac{\lambda_0 \bar{\cal E}e^{2i\phi}}{\sqrt{\epsilon}} \bigg[\cos(|\tilde \Omega^\prime| \hat t^2)\int_{-\infty}^{\hat t} \sin(|\tilde \Omega^\prime| s^2)ds\nonumber\\
&& \ \ \ \ \ \ \ -\sin(|\tilde\Omega^\prime|  \hat t^2)\int_{-\infty}^{\hat t} \cos(|\tilde \Omega^\prime|  s^2)ds\bigg]. \ \ \ \ \ \ \ \ \ \ \  \label{eq:bcres}
\end{eqnarray}
The ratio of mode and tidal forcing frequencies in the near-resonance region is
\begin{equation}
\tilde \Omega=\frac{\omega_0+\Delta \omega_0}{2\omega_{\rm orb}}
\end{equation}
and thus $\tilde \Omega^\prime=-3/8$ as in the nonspinning case since a modification of $\omega_0$ does not affect the derivatives. The asymptotic behavior of this solution away from the resonance is
\begin{equation}
\label{eq:inneraway}
\lim_{\omega_{\rm orb}\to \frac{\omega_0+\Delta \omega_0}{2}}Q_2^{\rm res}=-\frac{\lambda_0 \bar {\cal E} e^{2i( \tilde t_{\rm res}/\epsilon+\hat t/\sqrt{\epsilon}  )} }{2\sqrt{\epsilon}\, \hat t\, |\tilde \Omega^\prime|}+O(\hat t^{-3}, \epsilon^0).
\end{equation}
We see from the expansions \eqref{eq:outernearres} and \eqref{eq:inneraway} that the outer \eqref{eq:bcouter} and inner \eqref{eq:bcres} solutions match provided that we also introduce a shift in $\lambda_0$ in the near-resonance solutions given by
\begin{equation}
\lambda_0^{\rm near-res}=  \frac{\lambda_0}{1 + \Delta \omega_0 / \omega_0} . \label{eq:lambdashift}
\end{equation}
Note that as above we did not expand the denominator in $\Delta \omega_0$ in Eq.~\eqref{eq:lambdashift} to guarantee the matching. 
Finally, we can write down the composite solution for the quadrupole by combining the pre- and near-resonance solutions with the above modifications in each regime and subtracting their common singularity, as explained in Ref.~\cite{Steinhoff:2016rfi}. The last step is to ensure the correct limit at low frequencies $\omega_{\rm orb}/\Delta \omega_0 \ll 1$ by including an overall factor of $(1-\Delta\omega_0^2/\omega_0^2)$. 
We then compute the effective Love number $\Lambda_\ell=\lambda_{0\ell} \hat{k}^{\rm eff}_{\ell}/m_{1}^5$ used in the EOB code from
\begin{equation}
\hat k_{\rm eff} = -\frac{Q_{ij}E_{ij}}{\lambda_0 E_{kl} E_{kl}} = -\frac{\sum_m Q_{m}E_{m}}{\lambda_0 \sum_m E_{m} E_{m}} .
\end{equation}
We display the results here with the convention that the sign of $\Delta \omega_{0\ell}$ depends on the spin orientation and that it is a function of the EOB coordinate $r$ through $\omega_{\rm orb}=M^{1/2}r^{-3/2}$. The dynamical tidal enhancement factor including the shifts is then given by
\begin{align}
 \hat{k}^{\rm eff}_{\ell}&= a_\ell+b_\ell \left(1-\frac{(\Delta\omega_{0\ell})^2}{\omega_{0\ell}^2}\right)\bigg\{ \frac{\omega_{0\ell}^2 }{\omega_{0\ell}^2-(\ell\omega_{\rm orb}-\Delta \omega_{0\ell})^2}\nonumber\\
 &+\left[\frac{\omega_{f,\ell}^2}{2\sqrt{\epsilon_\ell}\, \hat t \, |\tilde\Omega^\prime| \left(1+\frac{\Delta \omega_{0\ell}}{\omega_{0\ell}}\right)(\ell \omega_{\rm orb})^2}\right]_{\omega_{f,\ell}= \omega_{0\ell}+\Delta \omega_{0\ell}}\nonumber\\
 &+\left[\frac{\omega_{f,\ell}^2}{\sqrt{\epsilon_\ell}(\ell\omega_{\rm orb})^2\left(1+\frac{\Delta \omega_{0\ell}}{\omega_{0\ell}}\right)} {\cal Q}_{\ell \ell}(\hat t)\right]_{\omega_{f,\ell}= \omega_{0\ell}+\Delta \omega_{0\ell}}\bigg\}, \ \ \  \label{keffDT}
\end{align}
where 
\begin{eqnarray}
{\cal Q}_{\ell\ell}(\hat t)&=&\cos(|\tilde\Omega^\prime| \hat t^2)\int_{-\infty}^{\hat t} \sin(|\tilde\Omega^\prime| s^2)ds\nonumber\\
&&-\sin(|\tilde\Omega^\prime| \hat t^2)\int_{-\infty}^{\hat t} \cos(|\tilde\Omega^\prime| s^2)ds.
\end{eqnarray}
 The quantities $\hat t(\omega_f)$ and the dimensionless parameter $\epsilon_\ell (\omega_{f})$ are given as explicit functions of $r$ by
\begin{equation}
\hat t=\frac{8}{5\sqrt{\epsilon_\ell}}\left(1-\frac{r^{5/2}\omega_{f,\ell}^{5/3}}{\ell^{5/3}M^{5/6}}\right),
\; \;
\epsilon_\ell=\frac{256 \mu  M^{2/3}\omega_{f,\ell}^{5/3}}{5\ell^{5/3}}.
\end{equation}
In Eq.~\eqref{keffDT} a body label on the quantities $\omega_{0\ell}$, $\hat t$, $\epsilon_\ell$, and 
${\cal Q}_{\ell \ell}$ is implied. 
For each $\ell$-multipole only $|m|=\ell$ contributes in Eq.~\eqref{keffDT} because the effect of modes with
$|m| \neq \ell$ has already been taken into account as adiabatic contributions.
For the quadrupole and octupole multipole moments the coefficients are given by 
$(a_2, a_3)=(\frac{1}{4},\frac{3}{8})$ and $(b_2,b_3)=(\frac{3}{4}, \frac{5}{8})$.

\section{Comparisons to numerical relativity simulations}
\label{sec:comparisons}

In this section, we compare the performance of the new extension 
of the \texttt{SEOBNRv4T} approximant derived in Sec.~\ref{sec:EOBmap} with numerical-relativity waveforms. 
We will focus on black hole -- neutron star (NSBH) systems simulated with the SpEC 
code~\cite{SPECWebsite,Duez:2008rb,Foucart:2012vn} and 
on binary neutron star (NSNS)  simulations with the
BAM code~\cite{Brugmann:2008zz,Thierfelder:2011yi,Dietrich:2015iva}.

\begin{figure}
\includegraphics[width=0.5\textwidth]{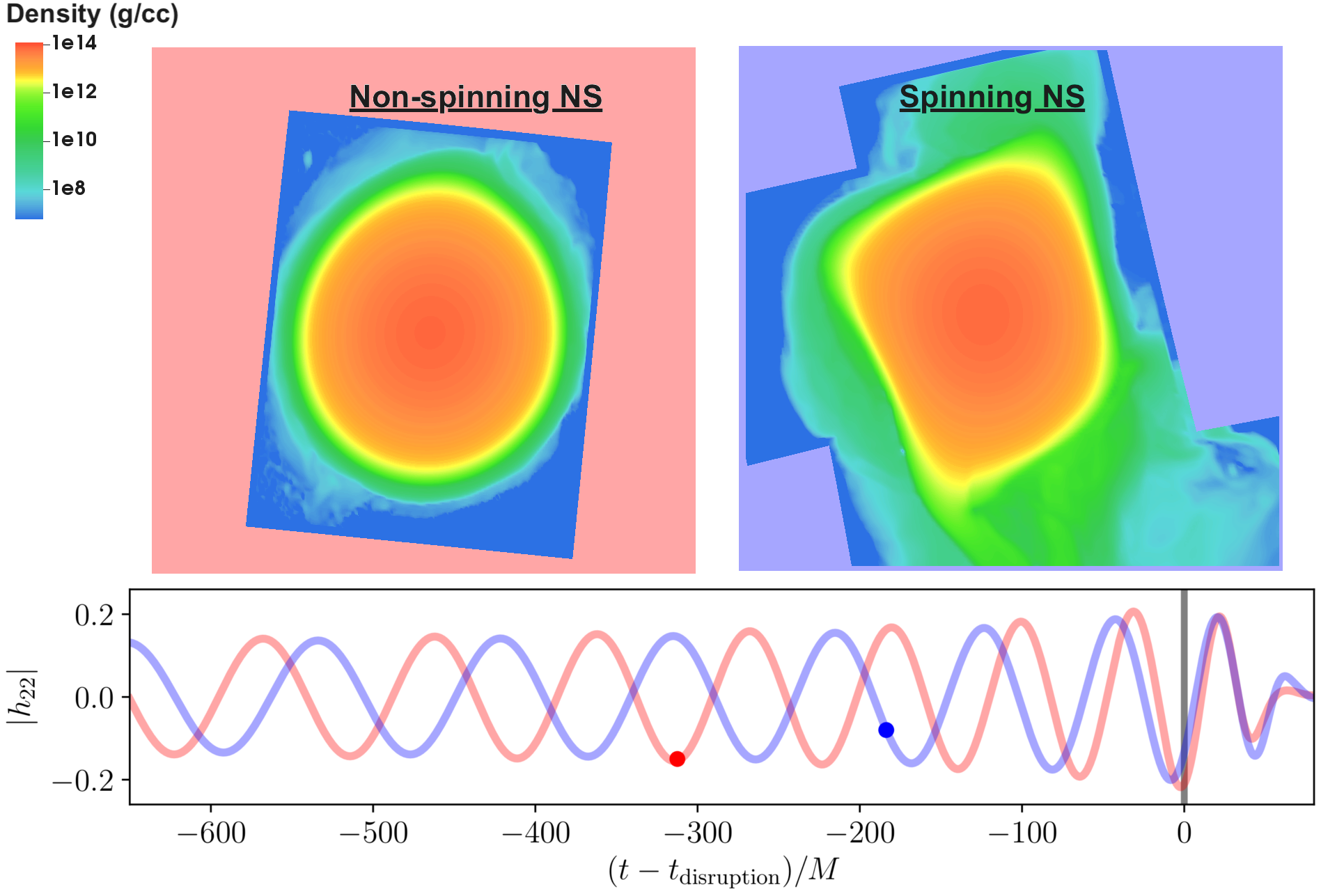}
\caption{\emph{Matter density in the late inspiral} for the non-spinning $q=1$ NSBH simulation ({\it top left}), 
and for the simulation with a spinning neutron star ({\it top right}), as well as the corresponding GW signal.  
We show the density in the equatorial plane at a time when the binary separation is $\sim 35\,{\rm km}$; cf.~markers in the bottom panel. 
The differences in the amount of distortion in the matter distributions (larger for the spinning NS) and onset of tidal disruption (earlier for the spinning NS), while being purely gauge-dependent, can be considered as a visualization of the enhancement of non-equilibrium tides.
The waveforms are aligned at the tidal disruption as determined from the peak GW amplitude.}
\label{fig:NSBHsims}
\end{figure}

In addition to the quantitative comparison that we will show in the next subsections, 
numerical-relativity simulations provide also 
a qualitative indication for the importance of non-equilibrium tides. 
Figure~\ref{fig:NSBHsims} shows a NSBH system with mass ratio $q=m_2/m_1=1$ and 
an anti-aligned neutron star spin, which enhances the excitation of non-equilibrium tides as discussed in Sec.~\ref{sec:response} and can be seen from Eq.~\eqref{keffDT} with a negative $\Delta \omega_{0\ell}$. 
While it is difficult to directly quantify these dynamical tides in a gauge-independent 
manner, the visual difference in Fig.~\ref{fig:NSBHsims} between the spinning and non-spinning 
matter distributions of the neutron star --  which have no invariant meaning -- provide an illustration of measurable differences in waveforms that we will analyze below.

\subsection{Comparison to NSBH SXS waveforms}
\begin{figure}
\includegraphics[width=0.48\textwidth]{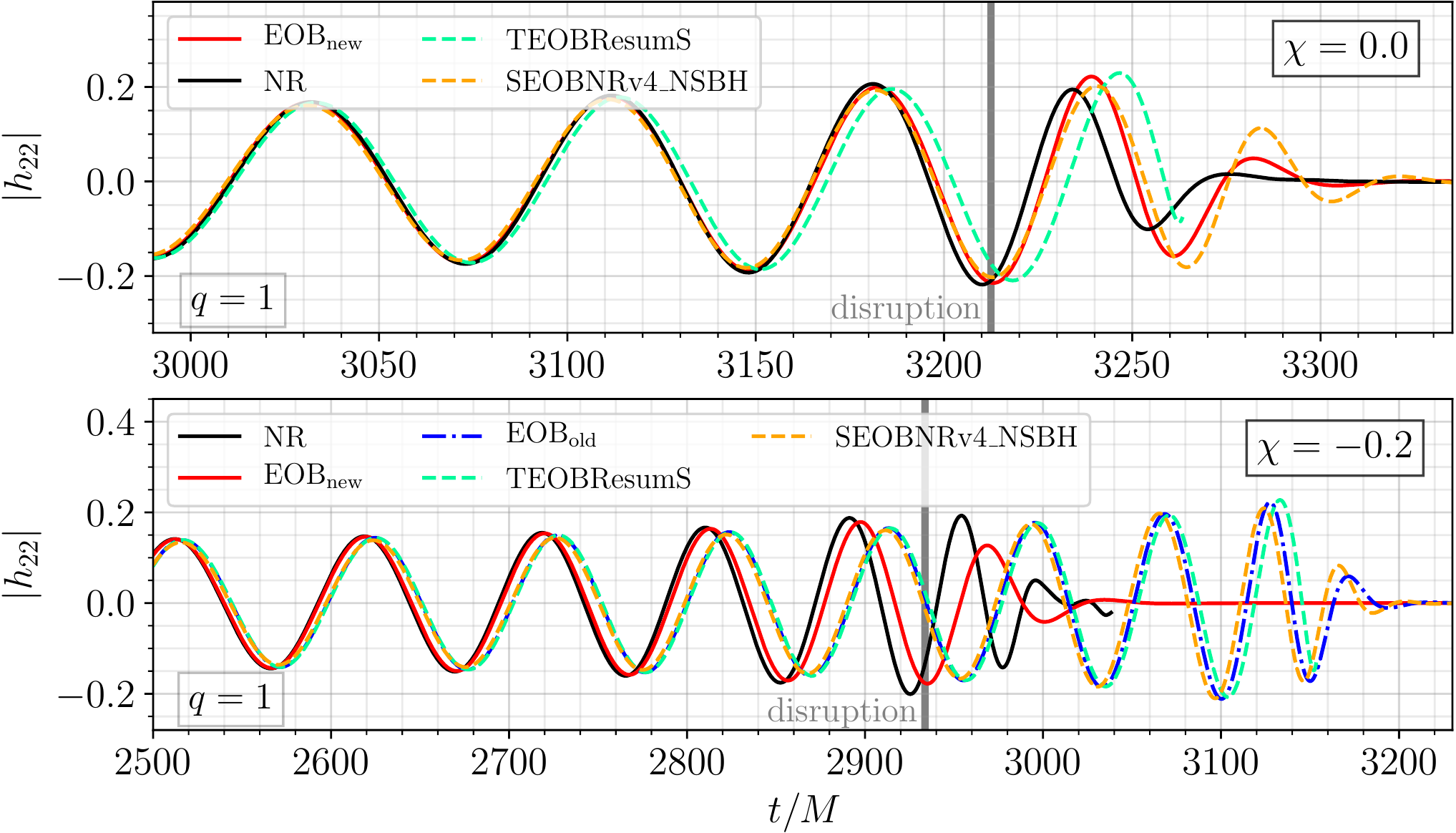}
\includegraphics[width=0.48\textwidth]{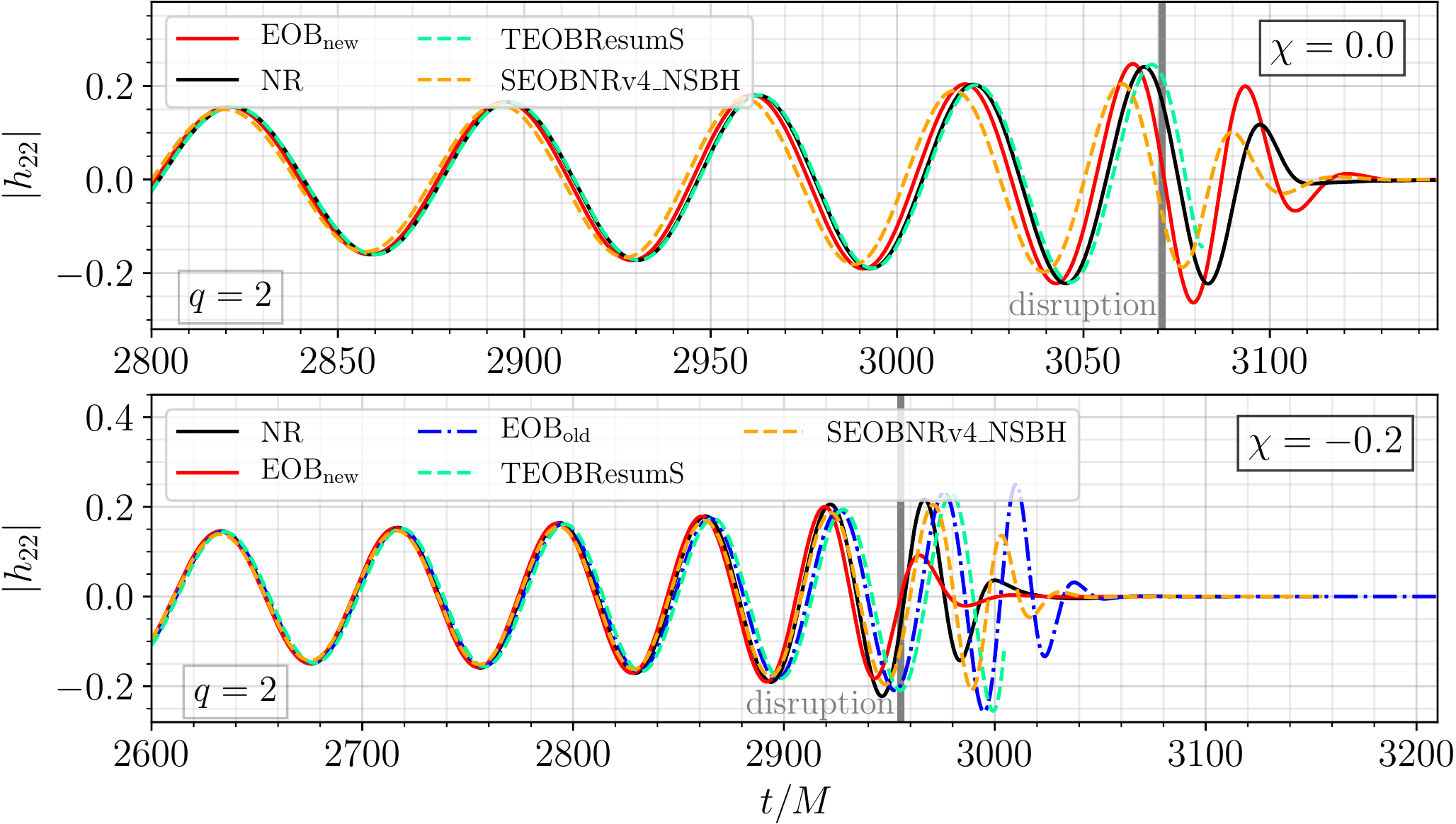}
\caption{\emph{Comparisons of various models to the NR waveforms for NSBH} considered in this work (top -- equal mass and bottom -- mass ratio $q=2$). The \texttt{TEOBResumS} waveforms are tapered to zero after the NSNS merger frequency, and the \texttt{SEOBNR} waveforms are tapered either at the NSNS merger or the $f$-mode resonance, whichever occurs first. Remarkably, the $f$-mode resonance in our new model including the spin shifts matches quite well with the tidal disruption frequency in the NR data, while the models that do not include the spin shifts predict a merger at a later time. 
\label{fig:comparison_SpEC}}
\end{figure}

For an initial comparison to numerical-relativity simulations and a validation of our new model, 
we consider two NSBH setups presented in Ref.~\cite{Foucart:2018lhe} and simulated with the SpEC code~\cite{SPECWebsite,Duez:2008rb,Foucart:2012vn}.
The two configurations represent an equal-mass $q=1$ and an unequal-mass $q=2$ setup employing a single polytropic equation of state $P=\kappa \rho^2$,
with $\kappa$ chosen so that $m_1/R_1=0.144$ (e.g.. $R_1=13.8\,{\rm km}$ if $m_1=1.35M_\odot$). Each mass ratio is simulated twice: with a non-spinning neutron star and with an anti-aligned dimensionless spin $\chi_1=-0.2$ on the neutron star. In both cases the black hole remains non-spinning. The neutron star spins and mass ratios were chosen as examples of a large expected impact of non-equilibrium tides. The numerical relativity data are publicly available in the SXS catalog~\cite{Boyle:2019kee}, where we use the simulations \texttt{SXS:BHNS:0004} and \texttt{SXS:BHNS:0005} for the $q=1$ and \texttt{SXS:BHNS:0002} and \texttt{SXS:BHNS:0007} for the $q=2$, nonspinning and spinning cases, respectively.
  
The evolutions start $10-13$ orbits before merger and use eccentricity-reduced ($e<0.002$), constraint satisfying initial conditions~\cite{Foucart:2008qt,Pfeiffer:2007yz}. Each case is simulated at three resolutions, and a detailed discussion of the estimated numerical error in these simulations can be found in Ref.~\cite{Foucart:2018lhe}. The same error estimates are used in this work. We typically find phase errors of less than $0.1\,{\rm rad}$ for most of the inspiral and rising to $(0.1-0.2)\,{\rm rad}$ at merger for the $q=1$ cases and to $(0.5-1.0)\,{\rm rad}$ for $q=2$.

We show in Fig.~\ref{fig:comparison_SpEC} the real part of the GW for the dominant (2,2)-mode for the two different mass ratios and different spin.
The parameters that we use in the EOB models were not obtained from quasi-universal relations [except for the fit in Eq.~\eqref{shiftfit}] and read
\begin{align}
  \frac{\lambda_0}{m_1^5} &= 799.3, & m_1 \omega_0 &= 0.06746, \\
  \frac{\Delta \omega_0}{\omega_0} &= -0.325, &  \frac{\lambda_{03}}{m_1^7} &= 2246, \\
  m_1 \omega_{03} &= 0.08805, & \frac{\Delta \omega_{03}}{\omega_{03}} &= -0.4, \\
  C_{ES^2} & = 7.14, & \frac{I}{m_1^3} &= 15.5 ,
\end{align}
recalling that these are the (quadrupolar, $\ell=2$) Love number $\lambda_0$, nonspinning $f$-mode frequency $\omega_0$, the mode shift $\Delta \omega_0$ due to spin, the corresponding octupolar ($\ell=3$) values $\lambda_{03}$, $\omega_{03}$, $\Delta \omega_{03}$, the dimensionless spin-induced quadrupole-moment constant $C_{ES^2}$ (normalized to 1 for black holes), and the moment of inertia $I$.
We see from these plots that for the nonspinning cases, existing waveform models predict the length of the waveforms and decrease of the GW amplitude due to the tidal disruption of the neutron star to a good approximation. However,  for systems with anti-aligned spins, all existing models, including those specialized for NSBH systems predict longer waveforms, while the new modification with spin effects (red curve) continues to yield a good prediction for the length of the signal. This is because in our new model, as also in the nonspinning \texttt{SEOBNRv4T} model, the GW signal is tapered to zero once the system reaches the $f$-mode resonance, if it occurs before the NSNS merger frequency. The fact that for the new \texttt{SEOBNRv4T} model the tapering occurs at the $f$-mode resonance for these cases can be seen by comparing to the \texttt{TEOBResumS} results, which are always tapered at the NSNS merger frequency. 
These findings highlight an interesting point, namely that there seems to be a direct relation between the $f$-mode resonance and the tidal disruption frequency.

 \begin{figure}
  \includegraphics[width=0.48\textwidth]{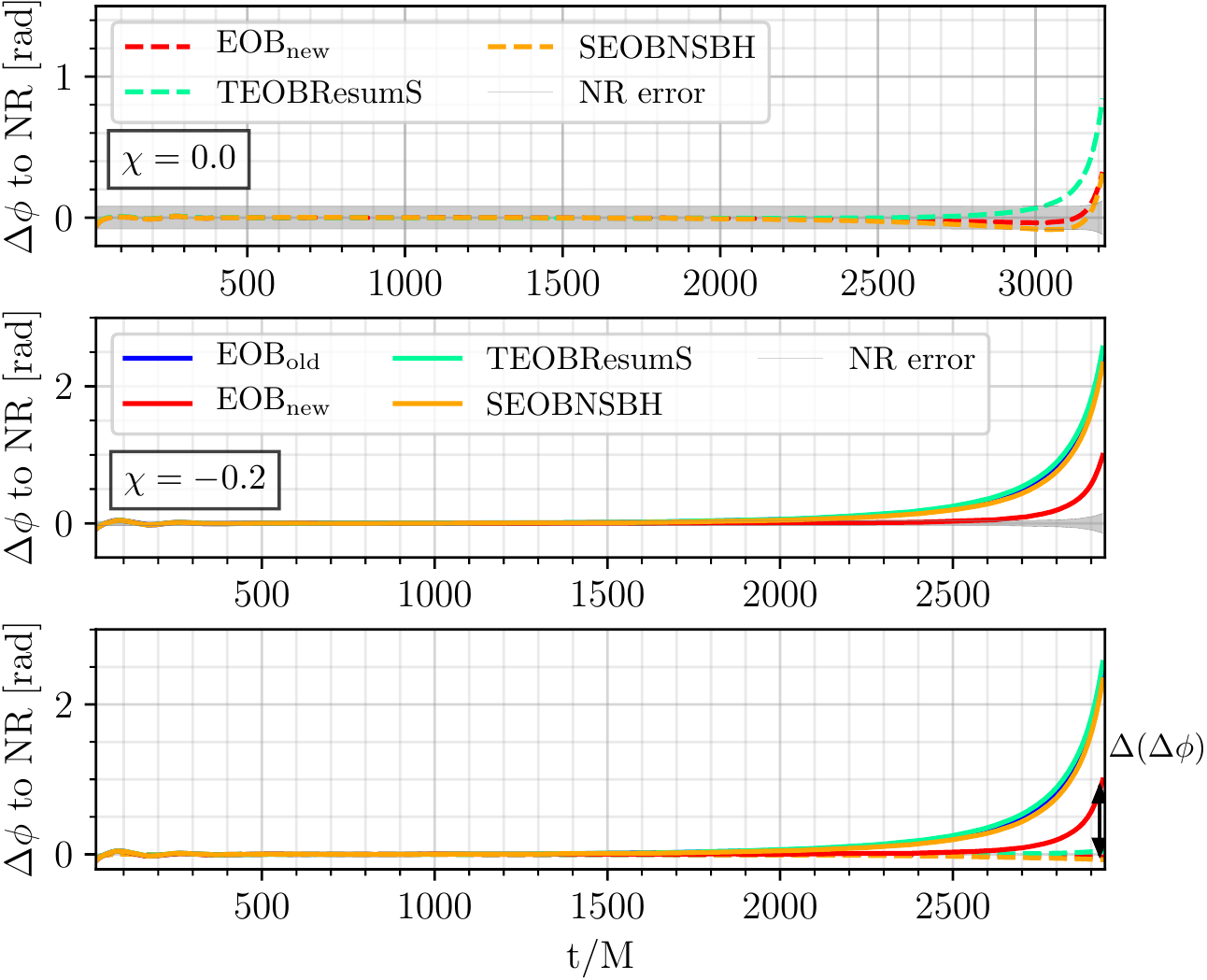}
  \caption{ 
  \emph{Phase differences in the NSBH case for different approximants and the numerical-relativity simulations}. 
  We consider the difference of the phase difference in the spinning minus the 
  nonspinning case for the $q=1$ NSBH simulation described in the main text
  to quantify systematic dependencies on the spin. 
  The top panel shows phase difference between the numerical-relativity simulation (and its uncertainty shown as a gray band) with the \texttt{TEOBResumS}, \texttt{SEOBNRv4T}, and SEOBNSBH models.
  The middle panel shows a similar comparison for an anti-aligned neutron star spin, where we also add the new \texttt{SEOBNRv4T} model
  described in this article. 
  In the bottom panel, we highlight again the phase difference between the approximants and the simulation results 
  in the spinning and non-spinning case. The difference between the phase difference, i.e., $\Delta(\Delta \phi)$, 
  is for the newly implemented model smallest.}
   \label{fig:comparison_SpEC_phase}
\end{figure}

Focusing on the $q=1$ setup, we next show the phase difference between the old (without the Coriolis effect) and the new \texttt{SEOBNRv4T} model, as well as results for \texttt{TEOBResumS}~\cite{Nagar:2018zoe} and \texttt{SEOBNSBH}~\cite{Matas:2020wab} in Fig.~\ref{fig:comparison_SpEC_phase}. 
For the non-spinning case (top panel) all models describe the GW phase accurately up to about 
one orbit before merger and stay within the estimated uncertainty of the numerical-relativity simulation 
(shown as the gray shaded region)~\cite{Foucart:2018lhe}. 
Considering the middle panel of Fig.~\ref{fig:comparison_SpEC_phase}, 
the anti-aligned spin of the neutron star enhances the dynamical tidal effects. We find that for this setup, the discrepancy in phasing between the new \texttt{SEOBNRv4T} model and the numerical relativity results is significantly less than for the other approximants. 
Therefore, we find a significantly better performance if non-equilibrium tidal effects are included.  
Although the new version of \texttt{SEOBNRv4T} is outside the estimated numerical uncertainty band close to the tidal disruption, 
other approximants show a noticeable dephasing even a few orbits before the disruption of the star, and it is this earlier-time regime where we expect to have more analytical control over the physics of the model. 

An even more important diagnostic of the robustness of our model, beyond a reduced phase difference in a few example cases, is that the 
phase differences to the numerical-relativity simulations are consistent between the spinning 
and non-spinning setups. To test the performance of our model with regards to this criterion we introduce the quantity $\Delta(\Delta)\phi$, which measures the phase difference for the case with spin minus the phase difference in the corresponding nonspinning case.  
A small $\Delta(\Delta)\phi$ indicates that the physics of the dynamical tidal effects and impact of the Coriolis effect are well-captured by the model, up to other physical effects with a different origin that are common among all cases. The definition of $\Delta(\Delta)\phi$ is illustrated in the bottom panel of Fig.~\ref{fig:comparison_SpEC_phase}
showing the phase difference between the EOB model and the SpEC data in the spinning case minus the phase difference for the non-spinning setup, i.e., the smaller the difference the better the consistency between the non-spinning and spinning data. 

 \begin{figure}
 \includegraphics[width=0.48\textwidth]{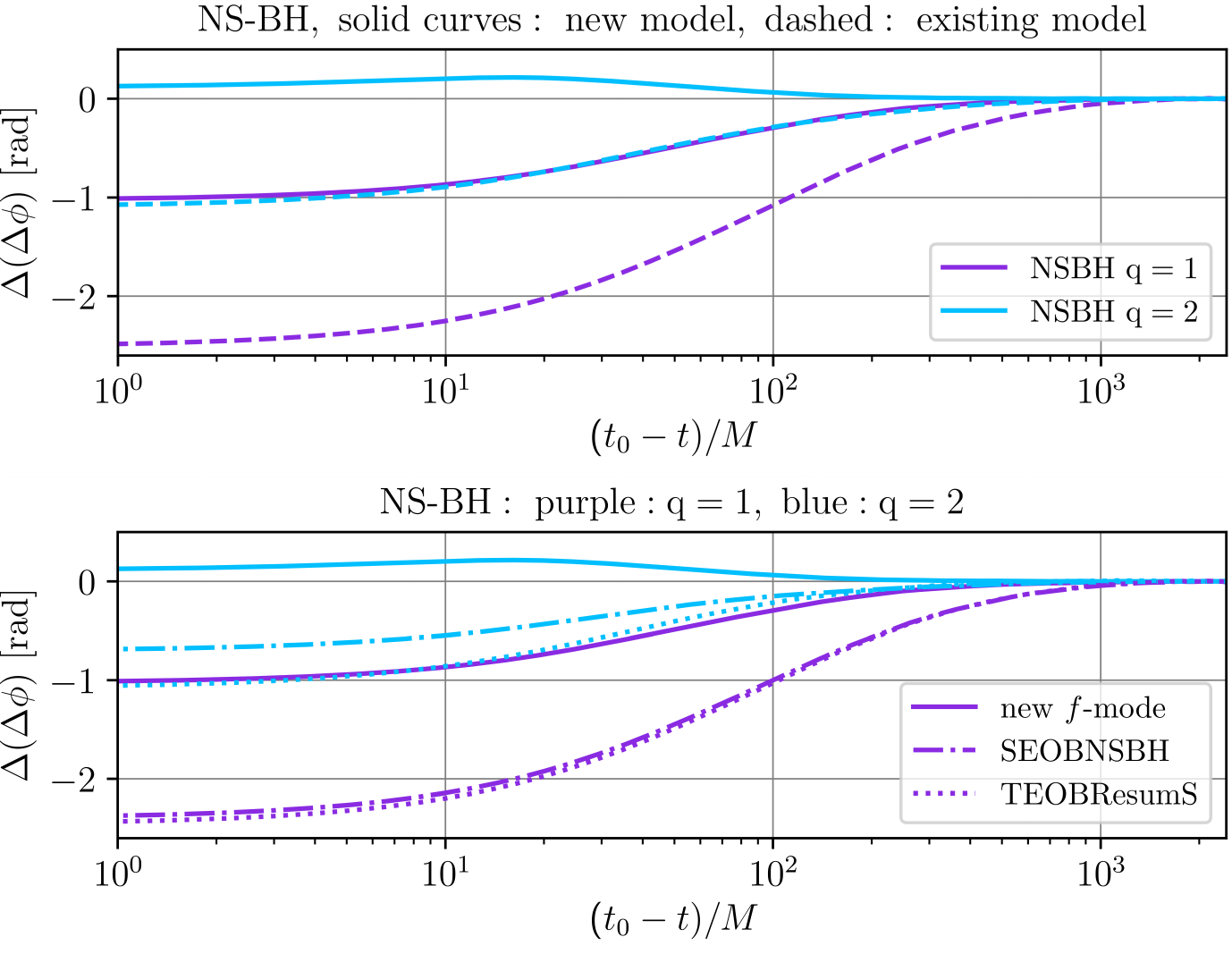}
 \caption{ 
 \emph{Phase differences compared to the phase differences in the nonspinning case for the NSBH configurations}. 
 Top panel: Solid curves are the new model developed in this paper, 
 dashed curves are the existing \texttt{SEOBNRv4T} model, 
 which does not account for spin effects on the mode resonances and exhibits 
 a wider spread in phase differences between spinning and non-spinning configurations. 
 Bottom panel: 
 Comparison with other waveform models which do not incorporate the spin-induced shift of the $f$-mode. We see that the new model has a consistently smaller spread. 
 In each case, $t_0$ is the time at which the first merger occurs 
 (i.e. for aligned spins it is the nonspinning merger, and otherwise the merger of the anti-aligned configuration). 
 }
\label{fig:comparison_deltadeltaphi_NSBH}
 \end{figure}

For a more quantitative presentation, 
we present $\Delta(\Delta \phi)$ for the new and old \texttt{SEOBNRv4T} model in the top panel of Fig.~\ref{fig:comparison_deltadeltaphi_NSBH} for the $q=1$ and $q=2$ setup. 
This clearly shows that for both configurations the new model outperforms the previous implementation. 
Similarly, the bottom panel shows also the new model in comparison to other NSBH approximants, 
where the disagreement between phase difference for spinning and non-spinning with respect to the corresponding 
numerical-relativity simulations is larger than for the model developed in this paper. 

\subsection{NSNS BAM waveforms}

\begin{table*}
  \centering
  \caption{NSNS-BAM configurations. 
    The first column defines the name of the configuration
    with the notation: EOS$_{m^A}^{\chi^A}$.
    The subsequent columns describe:
    the EOS~\cite{Read:2008iy}, 
    the NS' individual masses $m_{A,B}$, 
    the stars' dimensionless spins $\chi_{A,B}$,
    the Love number $\lambda_0$,
    the nonspinning $f$-mode frequency $\omega_0$,
    the mode shift $\Delta \omega_0$ due to spin,
    the corresponding octupolar ($\ell=3$) values $\lambda_{03}$, $\omega_{03}$, $\Delta \omega_{03}$,
    the dimensionless spin-induced quadrupole-moment constant $C_{ES^2}$ (normalized to 1 for black holes),
    and the moment of inertia $I$.
    The values here were not obtained from quasi-universal relations, except for the fit in Eq.~\eqref{shiftfit}.
    }     
\begin{tabular}{l|ccccccccccc}        
\hline
\hline
  Name & EOS & $m_{1,2} / M_\odot$ & $\chi_{1,2}$ & $\lambda_0 / m_1^5$ & $m_1 \omega_0$ & $\Delta \omega_0 / \omega_0$ & $\lambda_{03}/m_1^7$ & $m_1 \omega_{03}$ & $\Delta \omega_{03} / \omega_{03}$ & $C_{ES^2}$ & $I/m_1^3$ \\
     \hline
MS1b$_{1.35}^{-0.10}$ & MS1b & 1.3504 & $-0.099$ & 1528 & 0.05836 & $-0.137$ & 4488 & 0.07973 & $-0.162$ & 8.74 & 18.05 \\
MS1b$_{1.35}^{0.00}$  & MS1b & 1.3500 & $+0.000$ & 1528 & 0.05836 & $+0.000$ & 4488 & 0.07973 & $+0.000$ & --- & --- \\
MS1b$_{1.35}^{0.10}$  & MS1b & 1.3504 & $+0.099$ & 1528 & 0.05836 & $+0.145$ & 4488 & 0.07973 & $+0.164$ & 8.74 & 18.05 \\
MS1b$_{1.35}^{0.15}$  & MS1b & 1.3509 & $+0.149$ & 1525 & 0.05837 & $+0.210$ & 4474 & 0.07977 & $+0.238$ & 8.54 & 18.17 \\
\hline
 H4$_{1.37}^{0.00}$   & H4   & 1.3717 & $+0.000$ & 1003 & 0.06435 & $+0.000$ & 2605 & 0.08702 & $+0.000$ & --- & --- \\
 H4$_{1.37}^{0.14}$   & H4   & 1.3726 & $+0.141$ & 1003 & 0.06435 & $+0.202$ & 2605 & 0.08702 & $+0.232$ & 7.32 & 16.11 \\
\hline
SLy$_{1.35}^{0.00}$   & SLy  & 1.3500 & $+0.000$ & 389.6 & 0.07934 & $+0.000$ & 705.2 & 0.1067 & $+0.000$ & --- & --- \\     
SLy$_{1.35}^{0.05}$   & SLy  & 1.3502 & $+0.052$ & 389.6 & 0.07934 & $+0.084$ & 705.2 & 0.1067 & $+0.096$ & 6.18 & 12.34 \\     
SLy$_{1.35}^{0.11}$   & SLy  & 1.3506 & $+0.106$ & 388.8 & 0.07936 & $+0.164$ & 703.4 & 0.1068 & $+0.189$ & 5.59 & 12.41 \\ 
\hline      
\hline
     \end{tabular}
 \label{tab:BAM_data}
\end{table*}


 \begin{figure}
 \includegraphics[width=0.48\textwidth]{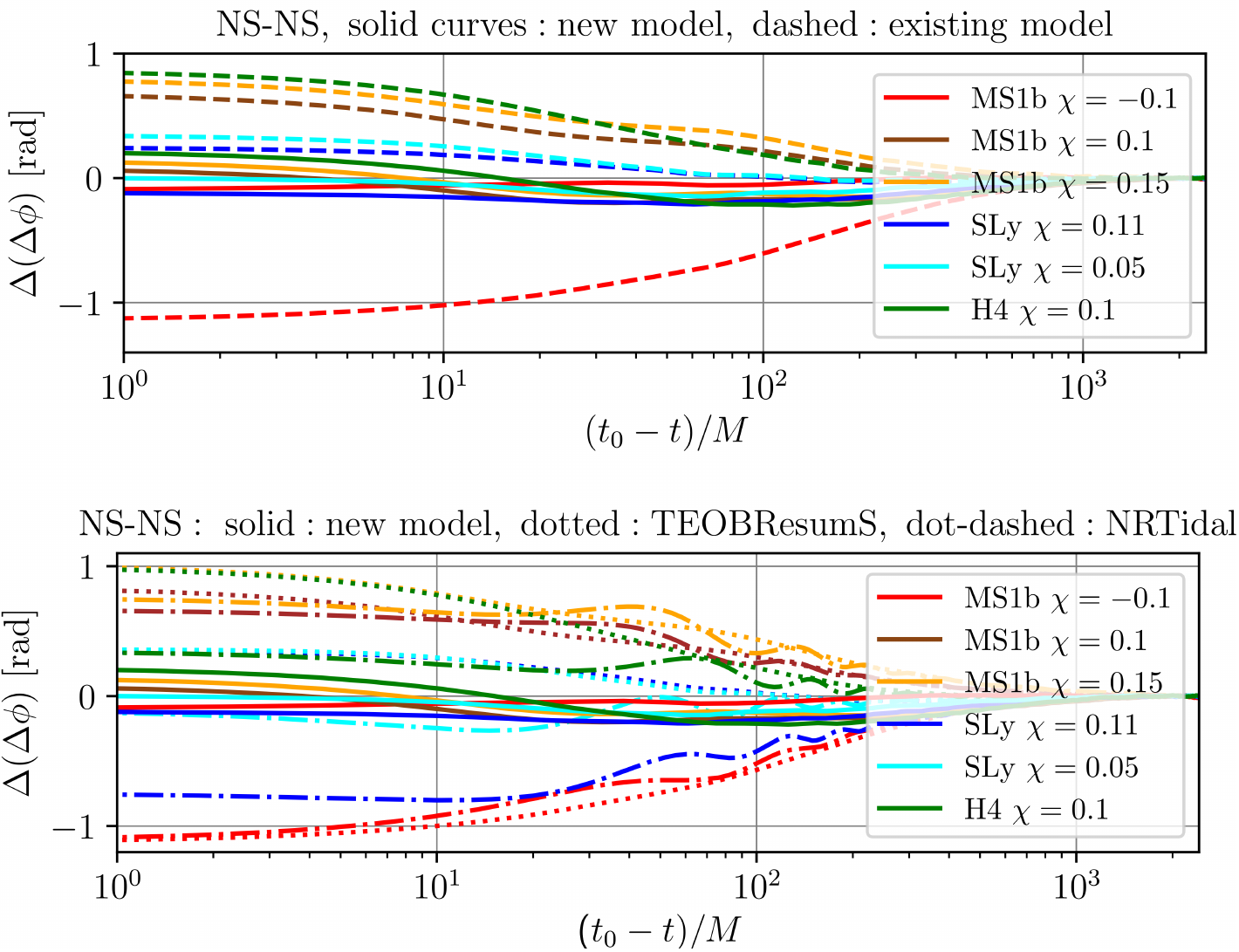}
 \caption{
 \emph{Phase differences compared to the phase differences in the nonspinning case for the NSNS configurations}. 
 Top panel: Solid curves are the new model developed in this paper, 
 dashed curves are the existing \texttt{SEOBNRv4T} model, 
 which does not account for spin effects on the mode resonances and exhibits 
 a wider spread in phase differences between spinning and non-spinning configurations. 
 Bottom panel: 
 Comparison with other waveform models which do not incorporate the spin-induced shift of the $f$-mode. We see that the new model has a consistently smaller spread. 
 In each case, $t_0$ is the time at which the first merger occurs 
 (i.e. for aligned spins it is the nonspinning merger, and otherwise the merger of the anti-aligned configuration). 
 }
 \label{fig:comparison_deltadeltaphi_BNS}
 \end{figure}

We continue our tests of the new model by comparing against numerical-relativity waveforms of NSNS systems computed with the BAM code~\cite{Dietrich:2017aum,Dietrich:2018upm,Dietrich:2018phi,Dietrich:2019kaq}, which include cases with both aligned and anti-aligned spins. 
In total, we consider three different equations of state: SLy, H4, MS1b. 
For all these equations of state, we consider one non-spinning configuration and one to three spinning setups; 
cf.~Tab.~\ref{tab:BAM_data} for further details and for the parameters used in the EOB models. The waveforms from Ref.~\cite{Dietrich:2018upm} show a clean second-order convergence, which allows using Richardson extrapolation to obtain a better guess for the true waveform, as 
discussed in Ref.~\cite{Bernuzzi:2016pie}. We use the
Richardson-extrapolated data from Ref.~\cite{Dietrich:2018upm} for our comparisons.

For all the cases, we follow a similar procedure as for the NSBH setups by 
focusing on the difference of the phase difference between the spinning EOB and the numerical-relativity waveforms 
with respect to their non-spinning counterparts. 
This way, we explicitly test the imprint of spin on the dynamical tides. 
Our results are summarized in Fig.~\ref{fig:comparison_deltadeltaphi_BNS}, 
where in the top panel the dashed lines refer to the old \texttt{SEOBNRv4T} model without the 
spin effects on the dynamical tides, and the solid lines show results for our new model. 
We find that for these cases, the new model shows a smaller phase difference between the numerical-relativity 
and the EOB data as the old model. We emphasize that this does not necessarily mean that the total 
phase difference with respect to the numerical-relativity data decreased in all cases, but rather that the phase difference for 
the non-spinning and spinning configurations becomes almost identical, indicating that the dependence on parameters is captured well.

The bottom panel of Fig.~\ref{fig:comparison_deltadeltaphi_BNS} compares the consistency of various 
GW models~\cite{Nagar:2018zoe,Dietrich:2019kaq} between the spinning and the non-spinning configurations for the NSNS binaries. 
We find that for all the setups the new \texttt{SEOBNRv4T} implementation has the smallest $\Delta(\Delta \phi)$, 
which means that the phase difference between the EOB model and the NR simulation is similar for the spinning 
and non-spinning cases. 
These results indicate that (i) the inclusion of spin-effects is consistent and (ii) further improvements
of the non-spinning sector will likely also improve the agreement between EOB and numerical-relativity predictions 
for spinning configurations. 

\section{Conclusions}
\label{sec:conclusion}

In this paper, we developed a ready-to-use waveform model that approximately captures the effects of spin on the $f$-mode dynamical tidal response of a neutron star. This model is based on the leading order terms in a relativistic effective action describing a spinning neutron star in a binary system, which we derived.

We found that within our approximation, a nonvanishing spin gives rise to a Coriolis interaction term in the action. We determined the coupling coefficient for this term from the spin-induced shift of the $f$-mode frequencies in slowly rotating relativistic neutron stars. A quasi-universal relation for this coupling coefficient was found as well, which is important for reducing the number of parameters to be inferred from GW observations. Further, using explicit post-Newtonian results we also analyzed relativistic effects (redshift and frame dragging) on the dynamical tidal response and found that they are subdominant compared to the spin effects.

We then developed a simple model that captures the main Coriolis effects on dynamical tides and incorporated it into a state-of-the-art EOB model. To test this new model, we performed comparisons to results from numerical relativity simulations of binary neutron star and neutron-star--black-hole binaries. The new model showed improved behavior over the parameter space compared to existing models that neglect the Coriolis effect. Moreover, we found that it predicted the tidal disruption frequency in mixed binaries significantly better than models without this spin-tidal effect. Our model is implemented in the LIGO Algorithms library. 

This work also identified important directions for future work. Moreover, the results from this paper provide a useful foundation for including these spin-tidal effects also in other waveform models. Improving the physics content of models is important for accurate measurements and robustness over a wider range in parameter space. We have also derived the relativistic effects on the response which can be included in future models, when we will also work out the effective Love number based on the spin-dependent response, allow for misaligned spins and other effects of the companion's spin, and also include other relativistic effects in a binary system.

\acknowledgements

We thank Gast\'on Creci for checking results in Sec.~\ref{sec:Newtaction}.
TH acknowledges support from the Nederlandse Organisatie voor Wetenschappelijk Onderzoek (NWO) sectorplan. FF acknowledges support from NASA through grant number 80NSSC18K0565, from the DOE through Early Career Award DE-SC0020435, and from the NSF through grant number PHY-1806278.
The authors are grateful for computational resources provided by the LIGO Laboratory and supported by the National Science Foundation Grants PHY-0757058 and PHY-0823459.

\appendix

\section{Connection to Ma, Yu, and Chen~\cite{Ma:2020rak}}\label{sec:compareMa}
\begin{figure}
\includegraphics[width=0.35\textwidth]{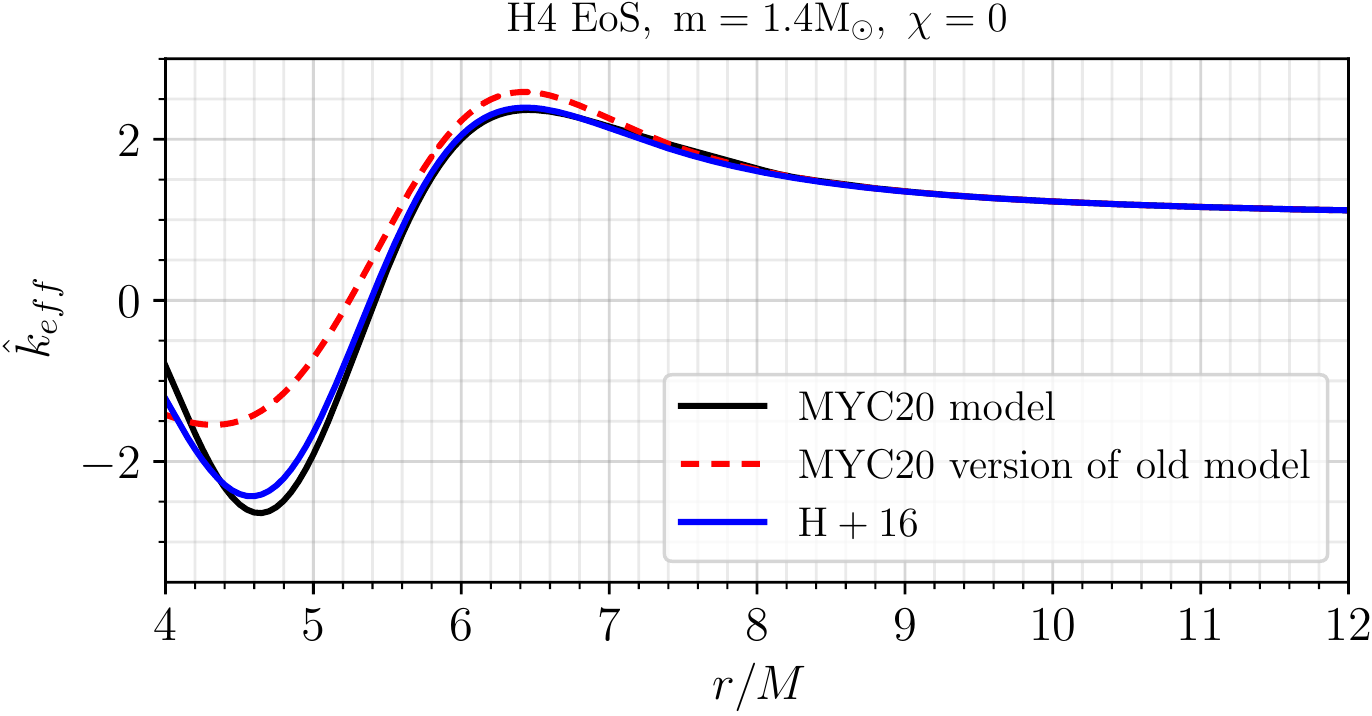}
\caption{\emph{Analytical approximations for a nonspinning equal-mass binary with one extended body}. Here, MYC20 is from Eq. (54) with the higher-order matching from Eq. (55) in Ref.~\cite{Ma:2020rak}. }
\label{fig:Ycompare_nospin}
\end{figure}

Here, we briefly outline similarities and differences with the work of Ma, Yu and Chen (hereafter MYC20) in Ref.~\cite{Ma:2020rak}, which considered the effect of spin on the $f$-mode in a Newtonian context. The effect of spins on the tidal response in Ref.~\cite{Ma:2020rak} is interconnected with the orbital dynamics obtained from numerical integrations of the equations of motion for the coupled system of dynamical quadrupoles, orbital variables, and gravitational radiation reaction. For ease of comparing to our results here, we apply the same procedure as for obtaining the effective Love number $\hat k_2^{\rm eff}$ oulined in Sec:~\ref{sec:EOBmap} to write the formulae in MYC20 explicitly as functions of the orbital separation $r$. The choice of this variable is motivated by the fact that the canonical coordinate $r$ plays a key role for the EOB dynamics, which are the basis of EOB waveforms. 
The formulae in Ref.~\cite{Ma:2020rak} are similar to our results, with slight differences in the definitions of variables such as $\hat t$ leading to small differences in the response near the resonance, as illustrated in Fig.~\ref{fig:Ycompare_nospin}. For instance, in our work, $\hat t$ is based on the phase and the parameter $\epsilon$, while in MYC20 $\hat t^Y$ is based on coordinate time and $\dot \omega$ at the resonance time and given explicitly by 
\begin{equation}
\hat t^{Y}=\frac{\sqrt{15}}{16 \, 2^{2/3}M^{1/3}\sqrt{\mu}\; w}-\frac{\sqrt{15} r^4 \; w}{128 \, 2^{1/3}M^{5/3}\sqrt{\mu}}
\end{equation}
where
\begin{equation}
w=\sqrt{-\omega_-\omega_++\Omega^2}-\Omega. 
\end{equation}
Here, $|\omega_{\pm}|$ are the frequencies of the two branches of $m$-modes whose frequencies coincide for $\Omega=0$.

\begin{figure}
\includegraphics[width=0.35\textwidth]{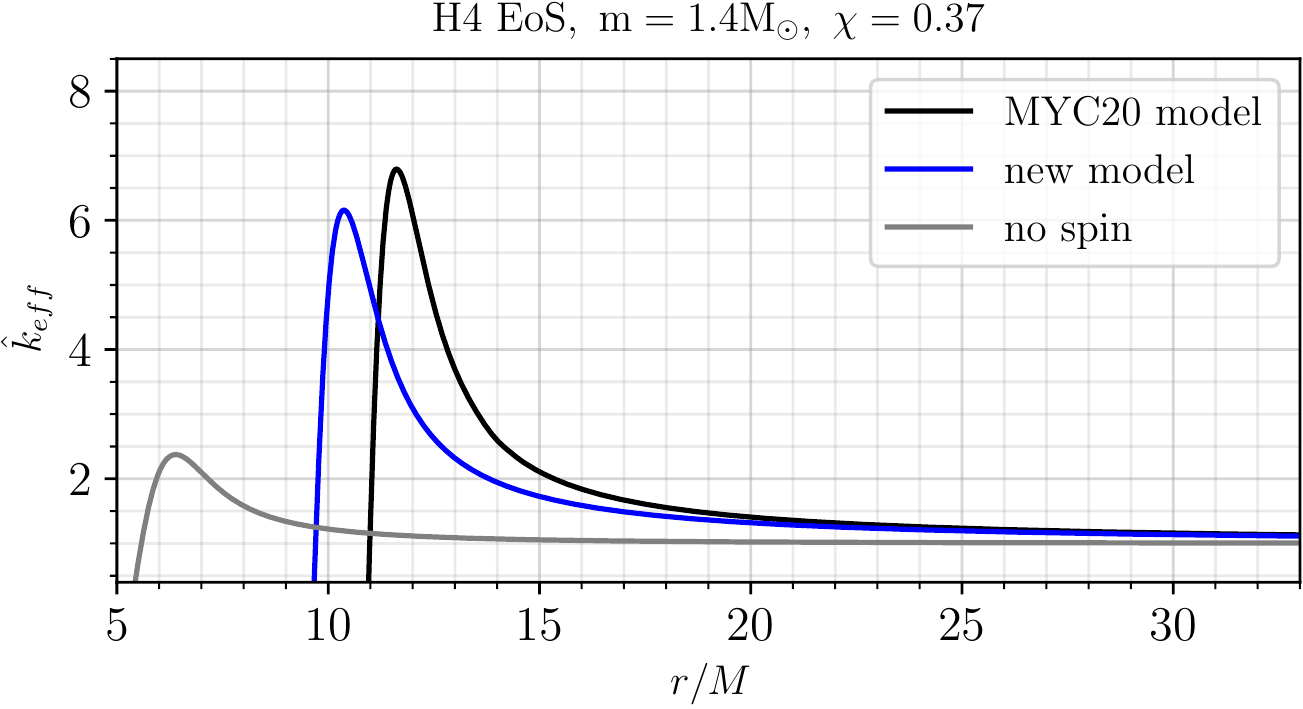}
\caption{\emph{Comparison between the model developed in this paper with $\{\Delta \omega_0,\Delta \lambda\}$ and MYC20 for a spin frequency $\Omega=2\pi\times 550$Hz}. For reference, the grey curve indicates the nonspinning result. }
\label{fig:Ycompare}
\end{figure}

MYC20 compute the $f$-mode frequencies for Newtonian Maclaurin spheroids. In the nonspinning case, this yields frequencies that are $\sim 320$Hz smaller than the relativistic values. MYC20 account for this by rescaling the density so as to match the relativistic frequencies. In this paper, we have used the fully relativistic results for the frequencies and their shifts, albeit only within the linear approximation for small $\Omega$. The variables of MYC20 are approximately related to these shifts by $|\omega_{\pm}|\approx \omega_0\mp \Delta \omega$. These different approximations and prescriptions also affect the orbital frequency at resonance. For instance, for the case of $\Omega=2\pi 550$Hz considered in MYC20, which is already outside of the linear regime, the resonance occurs at $\sim 300$Hz for MTY20 but not until $\sim 350$Hz for the parameters used here. The resulting predictions for the effective Love number are illustrated in Fig.~\ref{fig:Ycompare}.

\bibliography{nrcomparison,nrcomparison_auto}


\end{document}